\documentclass[lettersize,journal]{IEEEtran}
\usepackage{graphicx}
\usepackage{array}
\usepackage{balance}
\usepackage{booktabs}
\usepackage{url}
\usepackage{multirow}
\usepackage{subfloat}
\usepackage{comment}
\usepackage[linesnumbered,ruled,vlined]{algorithm2e}
\usepackage{caption}
\usepackage{orcidlink}
\usepackage{tikz}
\usepackage[dvipsnames]{xcolor}
\usepackage{gensymb}
\usepackage[numbers]{natbib}
\newcommand{\BAREAI}{\textbf{\texttt{BARE-AI}}}
\usepackage{amsmath}
\usepackage{amsfonts}
\usepackage{amssymb}
\usepackage{subcaption}
\usepackage{longtable}
\renewcommand\thesection{\Roman{section}}
\renewcommand\thesubsectiondis{\Alph{subsection}.}
\renewcommand\thesubsubsectiondis{\thesubsectiondis\arabic{subsubsection}}
\usepackage{lettrine}

\begin{document}
\title{\BAREAI: Bit-Flip Attack Resilience in AI Hardware through Built-in Performance Monitors
}

\author{{Habibur~Rahaman\orcidlink{0009-0000-2841-5572}},~\IEEEmembership{Student~Member,~IEEE},
        Swastik Bhattacharya\orcidlink{0000-0002-3146-8496},~\IEEEmembership{Student~Member,~IEEE},
        Sanjay Das\orcidlink{0000-0002-3146-8496},~\IEEEmembership{Student~Member,~IEEE},
        Kanad Basu\orcidlink{0000-0002-3146-8496},~\IEEEmembership{Senior~Member,~IEEE}
        ~Swarup~Bhunia\orcidlink{0000-0001-6082-6961},~\IEEEmembership{Fellow,~IEEE}
\thanks{H. Rahaman and S. Bhunia are with the Department of Electrical and Computer Engineering, University of Florida, Gainesville, FL, 32611, USA. (e-mail:rahaman.habibur@ufl.edu, swarup@ece.ufl.edu).\\
S. Bhattacharya and S. Das are with the Department of Electrical and Computer Engineering, University of Texas at Dallas, Richardson, TX, 75080, USA. (e-mail:SwastikBimal.Bhattacharya@UTDallas.edu, sanjay.das@UTDallas.edu).\\
K. Basu is with the Department of Electrical, Computer and Systems Engineering, Rensselaer Polytechnic Institute, Troy, NY, 12180, USA. (e-mail:basuk@rpi.edu).

The authors declare conflict of interest with all faculties from the University of Texas at Dallas, Rensselaer Polytechnic Institute and University of Florida.
}
}

\maketitle

\begin{abstract}
Deep Neural Networks (DNNs) are now integral to a wide range of safety critical systems, yet they remain highly vulnerable to bit-flip attacks (BFAs), in which a small number of memory level perturbations can drastically degrade model accuracy. Existing defenses often incur significant hardware overhead, depend on retraining, or fail against targeted adversarial flips. We propose \BAREAI, a runtime framework that detects, localizes, and mitigates BFAs during AI inference. \BAREAI~introduces AI Performance Counters (APCs), lightweight hardware monitors embedded in the accelerator datapath that capture fine grained per-layer activation statistics such as sparsity, entropy, kurtosis, and spectral shift. These statistics are analyzed by the \textit{Predictive Unit for Layer Security Evaluation (PULSE)}, a compact machine learning detector trained offline as an ensemble of classifiers and realized on-chip as a small neural engine. For explainability and recovery, \BAREAI~introduces an Activation Shift Index (ASI) for layer level fault localization and a z-score based statistical repair that resets anomalous weights toward clean layer statistics. Across CNNs, Vision Transformers, and Large Language Models under random, targeted, adaptive, and magnitude based BFAs, \BAREAI~achieves up to 98\% detection accuracy on vision models and between 74\% and 95\% on language models, restores near clean accuracy for CNNs and ViTs, and provides partial recovery for LLMs. Synthesized in a 28nm technology node, the monitoring infrastructure incurs less than 3\% energy overhead, less than 4\% area overhead, and approximately 10\% inference latency increase, with a configurable operating point that reduces latency overhead to about 6\%. A key property of the design is that its cost is independent of attack strength: unlike error correcting codes, whose redundancy grows with the number of tolerated flips, the monitoring overhead of \BAREAI~remains constant regardless of how many bits an adversary flips. These properties make \BAREAI~attractive for resource constrained, safety critical edge applications such as autonomous systems, energy, and healthcare.
\end{abstract}

\begin{IEEEkeywords}
DNN, AI Performance Counters, Bit-Flip Attack, AI Safety, AI accelerators, hardware security, runtime monitoring.
\end{IEEEkeywords}

\section{Introduction}
\begin{figure}[h]
    \centering
    \captionsetup{labelfont=bf}
    \includegraphics[width=0.40\textwidth]{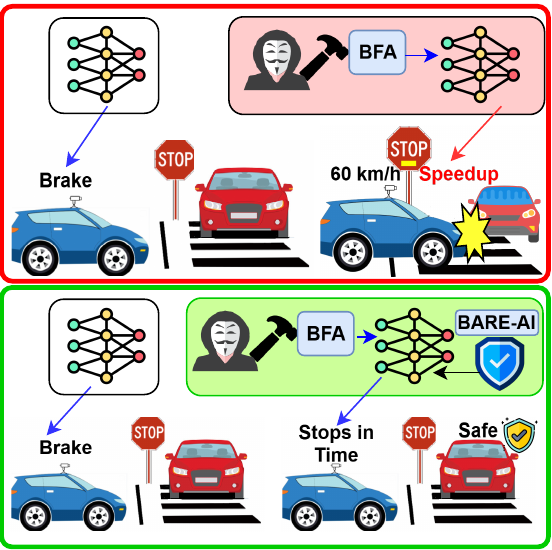}
\caption {Effect of a Bit-Flip Attack (BFA) on DNN models used in safety critical applications. \BAREAI~detects the attack at runtime, localizes the affected layers, and repairs corrupted weights using z-score based statistical correction.}
\label{fig:1}
\end{figure}

\IEEEPARstart{R}{apid} adoption of AI accelerators across emerging applications, spanning resource constrained edge devices to high performance cloud systems, has raised serious concerns about their security and reliability \cite{wang2023security, rahaman2024secure, rahaman2026evolving}. Although DNNs are widely deployed in safety critical settings, they are vulnerable to a broad class of adversarial manipulations. While much research has focused on input level threats such as evasion attacks, a more insidious threat lies in the manipulation of the model weights themselves. A particularly damaging class of such threats is the \textit{Bit-Flip Attack} (BFA), which corrupts parameters stored in memory \cite{rakin2019bitflip,liu2023neural, das2024attentionbreaker}. By flipping only a few bits in the model weights, an adversary can cause severe accuracy degradation or targeted misclassification~\cite{rakin2019bitflip}.
Figure~\ref{fig:1} illustrates the impact of a BFA in an autonomous driving system, where misclassification of a stop sign could trigger a catastrophic failure. For example, flipping just 13 bits out of 93 million in a quantized ResNet-18 reduced its top-1 accuracy from 69.8\% to 0.1\% \cite{rakin2019bitflip}. Large language models (LLMs) are similarly exposed: single digit numbers of bit flips can collapse the performance of billion parameter models such as Llama-3.2-1B-Instruct~\cite{das2024attentionbreaker}. Critically, these attacks leave no visible trace in the weight distribution; the model appears normal until the corruption manifests as failure.

Model parameters stored in DRAM or SRAM are vulnerable to diverse fault injection mechanisms, including Rowhammer induced charge leakage~\cite{gruss2023rowhammer}, laser injection~\cite{qian2023survey}, voltage glitching~\cite{rakin2019bitflip}, and radiation induced upsets~\cite{qian2023survey,willbold2025space}. Variants such as SledgeHammer~\cite{kang2024sledgehammer} and RowPress~\cite{luo2023rowpress} demonstrate that disturbance effects persist even in DDR4 systems, highlighting the fragility of DRAM~\cite{kim2014flipping,mutlu2020rowhammer}. DNN accelerator parameters held in DRAM are therefore susceptible to BFAs triggered by these physical fault injection methods \cite{kundu2024bit}, a threat intensified by nanoscale variability, reduced noise margins, and limited ECC protection~\cite{li2022survey}.
Attackers exploit these weaknesses with deliberate techniques such as voltage or clock faulting and electromagnetic interference~\cite{ji2023fault}, often combined with targeted search strategies like Progressive Bit Search~\cite{rakin2019bitflip,benedek2024impactful}. The threat is exacerbated in edge devices, which lack hardware redundancy, and in cloud settings, where Rowhammer can be launched from co-located virtual machines~\cite{frassetto2018imix}.
These attacks exploit an asymmetry of memory level faults: microscopic perturbations propagate across layers into catastrophic misbehavior, motivating lightweight runtime anomaly detection.

Existing defenses against BFAs differ in methodology, merits, and limitations.
Hardware methods such as ECC and redundancy mitigate random faults but fail against adversarial flips and incur substantial power, latency, and area cost \cite{benedek2024impactful,yao2020deephammer}. Training time defenses such as BFAT \cite{liu2023neural} improve robustness through weight corruption during retraining, but cannot detect or localize inference time faults and weaken under targeted or adaptive attacks. Software level methods, including weight randomization, memory layout obfuscation \cite{zhao2021fault}, and weight reconstruction \cite{li2020defending}, offer partial protection but require retraining, prior knowledge of attack locations, or system changes, limiting runtime applicability. Alternatives such as binarized networks \cite{siraj2021ra} improve resilience at the cost of accuracy and generality. Most existing defenses suffer from performance, scalability, or deployment limitations \cite{qian2023survey}.

\begin{figure}[h]
    \centering
    \captionsetup{labelfont=bf}
    \includegraphics[width=0.40\textwidth]{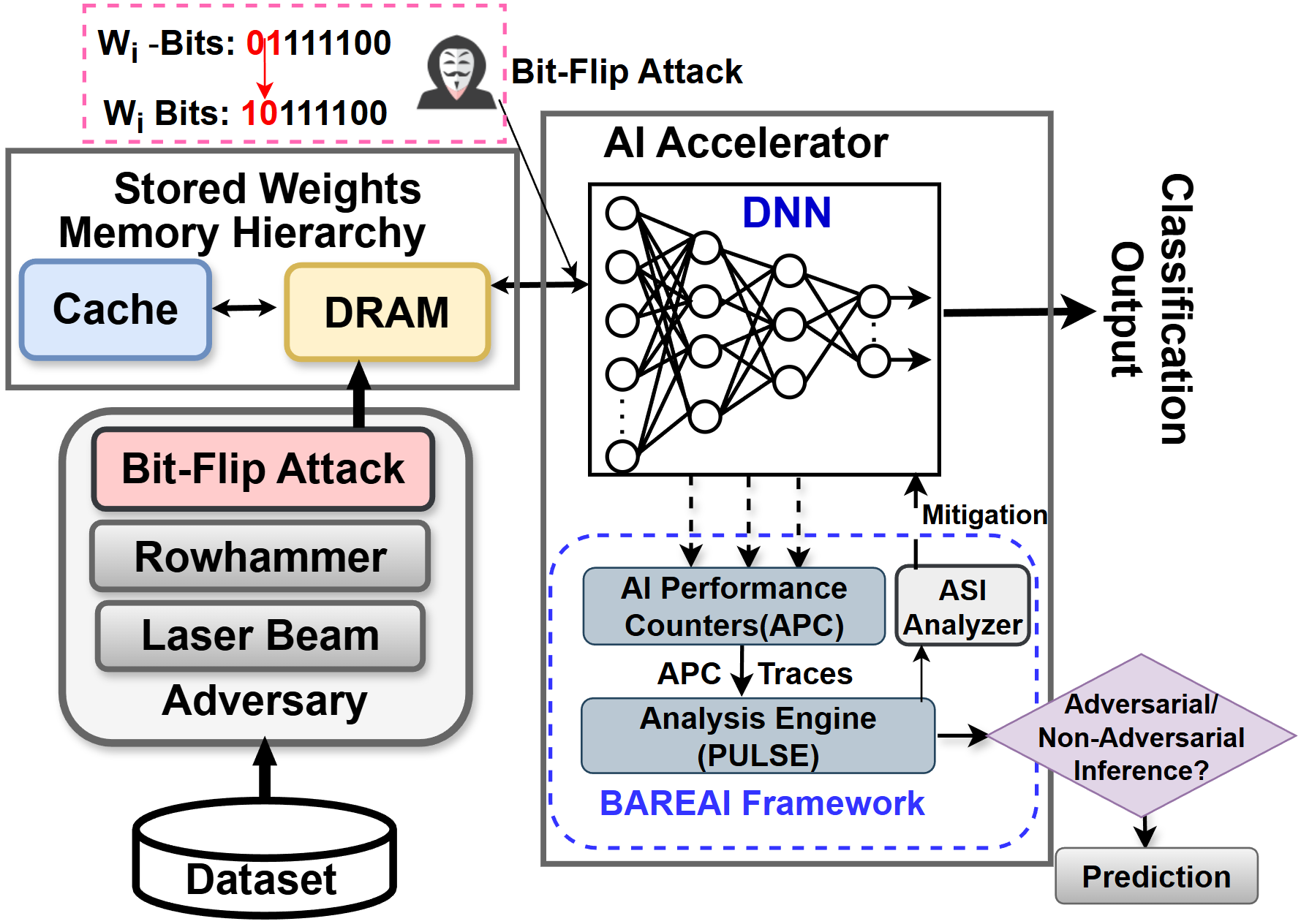}
\caption {\BAREAI~framework, showing how APCs and \textbf{PULSE} integrate into an AI accelerator. The APC captures operational traces, which are processed by \textbf{PULSE}; the ASI analyzer identifies the layers with the largest deviation and directs z-score based weight correction.}
\label{fig:21}
\end{figure}

Given these limitations, a practical defense should provide (i) runtime detection, (ii) attack agnostic applicability, and (iii) low computational overhead whose cost does not grow with attack strength. To address this, we propose \BAREAI, a lightweight runtime defense framework that performs behavioral monitoring of DNNs through dedicated hardware counters.
DNN accelerators depend on memory devices such as DDR4 for continuous parameter supply, tightly coupling their performance and security to memory integrity. Figure~\ref{fig:21} shows the block diagram of \BAREAI, which integrates an \textit{AI Performance Counter (APC)} block into an AI accelerator alongside an on-chip lightweight ML based analyzer called \textit{PULSE}. The APC records low level operational traces of the AI hardware, which PULSE processes for detection. As Figure~\ref{fig:21} indicates, DNN weights stored in DRAM are exposed to physical fault injection such as Rowhammer, laser, and electromagnetic interference \cite{kundu2024bit}.
Unlike weight inspection or retraining based methods, \BAREAI~tracks runtime statistics such as activation sparsity, entropy, kurtosis, and cosine drift to detect malicious deviations.

The key insight behind \BAREAI~is that, under normal conditions, each layer of a neural network exhibits statistically consistent activation patterns. Even a single bit flip that does not collapse the model perturbs these activation distributions and inter-layer dependencies. To capture such anomalies, \BAREAI~tracks the runtime metrics in Table~\ref{tab:layer_metrics} and constructs a behavioral signature of model execution. This signature is analyzed by PULSE, which distinguishes clean from tampered models. For explainability, \BAREAI~employs the Activation Shift Index (ASI) for layer level attribution of attack effects and applies a z-score based restoration mechanism to recover compromised performance where possible.

The paper makes the following major contributions:
\begin{itemize}
\item We present \BAREAI, a model agnostic, real time defense framework that unifies detection, localization, and mitigation of bit-flip attacks in a single hardware integrated pipeline, using activation level behavioral monitoring without weight inspection, retraining, or architectural modification of the DNN. In contrast to ECC based protection, whose redundancy grows linearly with the number of tolerated flips, the monitoring cost of \BAREAI~is constant and independent of attack strength (Section~\ref{sec:discussion}).
\item We formalize AI Performance Counters (APCs)~\cite{rahaman2026samurai}, a set of per-layer activation statistics realized as a configurable hardware counter block inside the accelerator datapath, and present the PULSE detection methodology, in which an offline trained ensemble is realized on-chip as a compact neural engine.
\item We develop the Activation Shift Index (ASI) for layer level fault localization and a z-score based mitigation technique that resets anomalous weights toward clean layer statistics while preserving benign parameters.
\item We evaluate \BAREAI~across CNNs, Vision Transformers, and LLMs under random, targeted, adaptive, and magnitude based BFAs, including INT8 quantized settings, and characterize hardware cost through 28nm synthesis. We report both the strengths of the approach (near clean recovery for CNNs and ViTs, high detection rates) and its current limits (partial recovery for deep language models).
\end{itemize}

The remainder of the paper is structured as follows. Section II covers preliminaries and related work. Section III presents the methodology, including the threat model for the monitoring infrastructure itself. Section IV presents the experimental findings. Section V discusses \BAREAI~in relation to existing defenses, and Section VI concludes the paper.

\section{\textbf{Background and Related Work}}
In this section, we present the threat model for BFAs and review the defenses that have been proposed.

\subsection{\textbf{Threat Model and Attack Surface}}
\label{sec:threat_model}
To assess the capability of the proposed defense framework against corruption of the inference time model, we first define the class of adversarial behaviors against which the system must defend. We adopt a threat model in which adversaries compromise the integrity of deployed DNNs by inducing parameter perturbations, specifically bit flips, in stored weight parameters. The model weights, encoded in FP32, FP16/BF16, or INT8 formats, may reside in off-chip DRAM or on-chip SRAM or cache. Adversaries can exploit gradient guided or heuristic search techniques such as Progressive Bit Search (PBS)~\cite{rakin2019bitflip, das2024attentionbreaker, kundu2024bit} and Targeted Bit-Flip Attacks (TBFA)~\cite{liu2023neural} to identify highly sensitive weight bits for maximal disruption. Realistic fault injection mechanisms include disturbance induced memory corruption such as Rowhammer~\cite{gruss2023rowhammer}, targeted neural network specific campaigns such as DeepHammer~\cite{yao2020deephammer,benedek2024impactful,kundu2024bit}, and broader techniques involving voltage or clock glitching, electromagnetic or laser injection, and software based stressors~\cite{ji2023fault}. The resulting perturbations may manifest as isolated single bit errors, clustered flips within a codeword, or burst faults across adjacent memory locations, patterns that can evade simple error correcting codes or overwhelm hardware scrubbing.
Whether such perturbations arise from malicious attacks, including parasitic trojans in the ML infrastructure~\cite{rahaman2026ispy}, or from non-malicious hardware faults, they introduce critical errors into the computational flow, propagate across layers, and ultimately compromise model integrity, an unacceptable risk in safety critical applications. Our objective is to reliably detect \textbf{any deviation} from expected model behavior and enable timely mitigation. The security of the monitoring infrastructure itself, and the reason it does not share the attack surface of the model weights, is analyzed in Section~\ref{sec:monitor_security}.

\subsection{\textbf{Defenses of Bit-Flip Attacks (BFAs)}}
Existing defenses against BFAs can be categorized by the stage at which protection is applied: hardware, training time, system (runtime), and software. Hardware level approaches include error correcting codes (ECC), memory scrambling or layout randomization, and dual modular redundancy (DMR). ECC corrects random faults but incurs area and latency overhead and is ineffective against targeted adversarial flips \cite{li2020defending}. Memory scrambling and layout randomization reduce attacker precision but can be reverse engineered via side channels, while DMR detects faults through redundant execution at the cost of doubled computation.
Training time defenses aim to improve robustness during model optimization. Techniques such as Bit-Flip Adversarial Training (BFAT) inject bit flips during retraining \cite{liu2023neural}, neural fingerprinting detects deviations from clean activation patterns, and activation constraining methods limit fault propagation through range restriction or boundary aware clipping \cite{chen2021low, zhan2021improving}. Fault aware training and range aware quantization further improve resilience by constraining vulnerable weights \cite{itc2024_fatrabbit, tmlr2023_quantization}. Although effective, these methods depend on offline retraining, are architecture specific, and do not provide runtime fault detection or recovery. System level monitoring using hardware performance counters (HPCs) captures execution anomalies \cite{chenet2024hardware}, while software level techniques include robust quantized models \cite{siraj2021ra}, selective weight protection \cite{wang2302aegis}, and runtime detection and correction mechanisms such as NeuroPots \cite{liu2023neuropots}. These approaches remain limited by hardware dependence, static configurations, or insufficient fault localization. Meanwhile, attacks such as DeepHammer \cite{yao2020deephammer}, impactful search \cite{benedek2024impactful}, and quantized attacks \cite{kundu2024bit} continue to grow more efficient, motivating stronger defenses.

\textbf{Positioning of this work.} Individual ingredients of our approach have appeared in prior contexts: HPC based anomaly detection monitors processor level events for malware and integrity violations \cite{chenet2024hardware}, activation fingerprinting compares runtime activations against reference patterns \cite{liu2023neural}, statistical bounds on weights have been used for reconstruction based repair \cite{li2020defending}, and NeuroPots performs runtime detection with honeypot neurons \cite{liu2023neuropots}. \BAREAI~differs from these lines of work in four respects. First, the monitored signal is a purpose built set of per-layer activation statistics captured by dedicated counters inside the accelerator datapath, rather than general purpose CPU event counters that observe the workload only indirectly. Second, detection, layer level localization, and statistical repair are integrated into a single runtime pipeline; prior activation fingerprinting detects anomalies without localizing or repairing them, and reconstruction based repair \cite{li2020defending} requires retraining and prior fault identification. Third, the hardware cost of the monitor is independent of the number of adversarial flips, in contrast to ECC whose redundancy grows with the required correction strength (Section~\ref{sec:discussion}). Fourth, the same monitoring infrastructure applies unchanged across CNNs, Vision Transformers, and LLMs, since the counters observe generic activation statistics rather than architecture specific structures. We view the contribution as the co-designed system and its hardware realization, together with the ASI localization metric and the evidence that activation level counters retain discriminative power under quantization, adaptive attacks, and cross domain transfer.

To address the gaps above, we propose \BAREAI, a model agnostic runtime behavioral monitoring framework that provides real time detection, localization, and recovery from bit flip induced distribution shifts. Using a lightweight hardware assisted mechanism that tracks activation statistics via APCs, \BAREAI~delivers fine grained explainability and robust protection with low overhead across CNN and Transformer architectures.

\subsection{\textbf{Model Quantization}}
Quantization reduces the computational and memory overhead of deep neural networks, enabling deployment on resource constrained hardware such as consumer grade GPUs or edge devices. Converting floating point weights and activations to low bit formats such as INT8 significantly lowers memory and bandwidth requirements with minimal architectural changes. Well designed quantization schemes (uniform or per-channel) preserve accuracy and reduce inference latency even with aggressive bit width reductions \cite{jacob2018quantization, banner2019post, chen2022clipping}. Post-training quantization (PTQ) is especially practical, as it operates on pretrained models without retraining or access to training data \cite{jacob2018quantization, lee2022quantune}. For large architectures such as LLMs, quantization methods have evolved to meet precision and stability demands. Toolkits such as BitsAndBytes enable INT8 post-training quantization of LLMs using mixed precision and vector wise strategies. The LLM.int8() method \cite{dettmers2022gpt3} applies 8-bit computation to most weights while isolating rare large magnitude outliers for 16-bit processing, allowing over 99.9\% of matrix operations to run in INT8. This hybrid approach significantly reduces inference memory while preserving near full precision accuracy for large models \cite{dettmers2022gpt3, xiao2023smoothquant}. In this work, we adopt BitsAndBytes based INT8 PTQ to evaluate the performance of \BAREAI~against BFAs on quantized models.

\begin{table}[t]
\Large
\centering
\captionsetup{labelfont=bf}
\caption{AI accelerator metrics extracted by APC} 
\renewcommand{\arraystretch}{1.2} 
\resizebox{\columnwidth}{!}{ 
\begin{tabular}{|l|p{12cm}|}
\hline
\textbf{Metric} & \textbf{Description and Equation} \\ \hline
\textbf{Layer Sparsity} & Ratio of zero-value elements in a layer's activation. \\
& \( \boldsymbol{S_L = \frac{\#\{o_i \in O | o_i = 0\}}{\#\{o_i \in O\}}} \), where \( \boldsymbol{o_i} \) is the \( i \)-th element in the output tensor \( \boldsymbol{O} \), and \( \boldsymbol{\#\{\cdot\}} \) denotes the cardinality of the set, \( \boldsymbol{L} \) denotes the neural network layer with output tensor
\(\boldsymbol{O}\), \(\boldsymbol{S_L}\) is the layer sparsity. \\ \hline
\textbf{Sparsity} &  Fraction of near-zero activations.
    \[
    S_l = \frac{1}{Bd} \sum_{i=1}^{B} \sum_{j=1}^{d} \mathbf{1}(|A_l^{(i,j)}| < \epsilon)
    \] \\ \hline

\textbf{Layer Zero Counts} & It is the absolute count of zero-
valued elements within each layer's activation. \\
& \( \boldsymbol{Z_L = \sum_{i=1}^{N} \delta(o_i = 0)} \) \\
&  \( \boldsymbol{N} \) is the total number of elements in \( \boldsymbol{O} \), \( \boldsymbol{o_i} \) is the \( i \)-th element of \( \boldsymbol{O} \), and \( \boldsymbol{\delta(\cdot)} \) is the indicator function. \\ \hline

\textbf{Dense Layer Activities} & Includes average activation, sparsity, and extreme values. \\
& \textbf{Average Activation:} \( \boldsymbol{\bar{a}_L = \frac{1}{N} \sum_{i=1}^{N} \sum_{j=1}^{M} a_{ij}} \) \\
& \textbf{Sparsity:} \( \boldsymbol{S_L = \frac{\#\{a_{ij}\in A | a_{ij} = 0\}}{\#\{a_{ij}\in A\}}} \) \\
& \textbf{Max/Min Activation:} \( \boldsymbol{a_{\max} = \max(A), a_{\min} = \min(A)} \) \\
&  \( \boldsymbol{\bar{a}_L} \) is the average activation, \( \boldsymbol{S_L} \) is the sparsity, \( \boldsymbol{a_{\max}} \) and \( \boldsymbol{a_{\min}} \) are the maximum and minimum activations, \( \boldsymbol{A} \) is the activation matrix, \( \boldsymbol{N} \) is the number of inputs, and \( \boldsymbol{M} \) is the number of neurons in the dense layer. \\ \hline

\textbf{FLOPs per Layer} & Total floating-point computations per layer. \\ \hline

\textbf{Layer-wise TOPs} & Tera Operations Per Second per layer, scaled FLOPs. \\ \hline

\textbf{Layer-wise MACs} & Multiply-Accumulate operations per output element. \\ \hline

\textbf{Layer-wise FLOPs} & Individual FLOPs assessment per layer. \\ \hline

\textbf{Layer-wise Entropies} & Shannon entropy of the normalized activation distribution.
    \[
    H_l = - \sum_{k} p_k \log p_k \quad \text{where } p_k = \frac{n_k}{\sum_k n_k}
    \] \\ \hline

\textbf{Throughput} & Inverse of inference time; images processed per second. \\ \hline
    \textbf{Activation Energy} & $E_\ell = \sum_{i,j} (A_\ell^{(i,j)})^2$ \\ \hline
\textbf{Kurtosis ($K_l$)} & Measures the heaviness of the tail of activations.
    \[
    K_l = \frac{\mathbb{E}[(A_l - \mu)^4]}{\sigma^4}
    \] \\ \hline
\textbf{Cosine Drift ($C_l$)} & Drift in cosine similarity between the stored clean reference activation profile and the current inference:
    \[
    C_l = 1 - \cos(\mathbf{A}_l^{\text{clean}}, \mathbf{A}_l^{\text{test}})
    \] \\ \hline

\textbf{FFT Spectrum Shift ($F_l$)} & Frequency-domain shift in activations via Fast Fourier Transform.
\\ \hline
\end{tabular}
}
\label{tab:layer_metrics}
\end{table}

\section{\textbf{Methodology}}
In this section, we present the design of \BAREAI, a lightweight runtime defense framework that leverages layer wise behavioral signals, termed \textit{AI Performance Counter (APC)} metrics, to detect and recover from bit-flip attacks. The framework does not require retraining of the protected model.
\vspace{-2mm}
\subsection{\BAREAI~Framework}
\begin{figure*}[h]
    \centering
    \captionsetup{labelfont=bf}
    \includegraphics[width=0.7\textwidth]{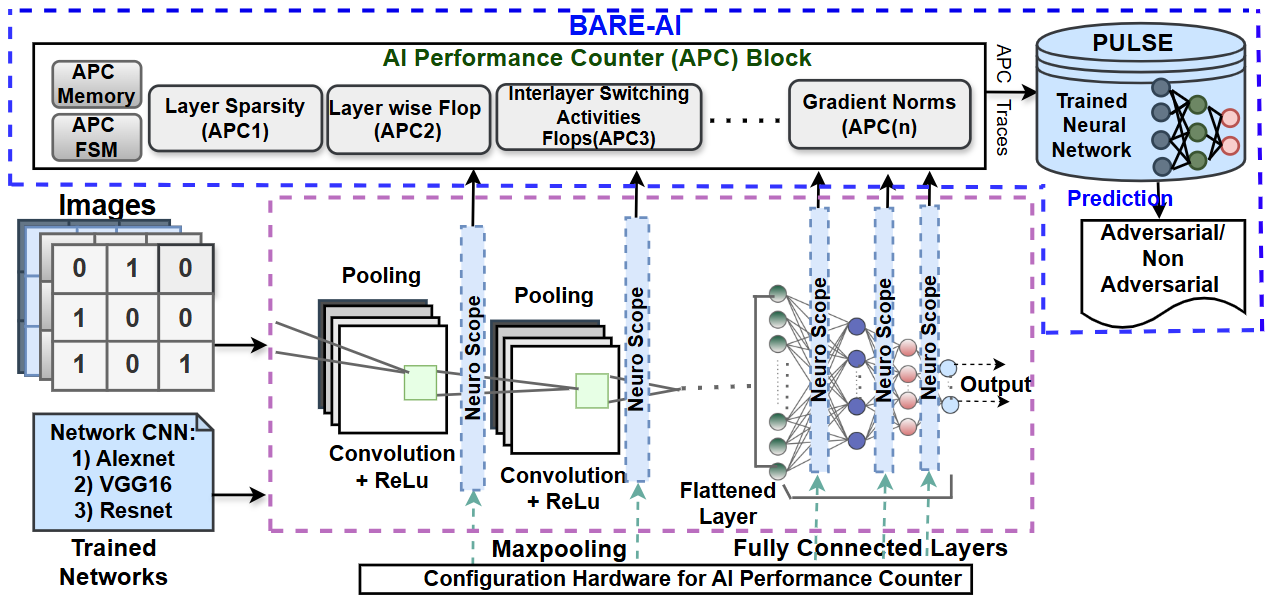}
    \caption {
    AI Performance Counter (APC) in an AI accelerator, illustrating the architecture of the APC module, which consists of a Finite State Machine (FSM) and the APC memory block that records operational metrics during AI inference and transmits the data to \textit{PULSE} for real-time bit-flip attack detection.
        }
    \label{LWM}
\end{figure*}

\BAREAI~is integrated with the AI accelerator as shown in Fig.~\ref{LWM}. It is composed of four major operations: i) \textit{APC Extraction}, which captures a rich set of activation based metrics across layers during inference; ii) the \textit{PULSE Classifier}, which determines whether the model has been compromised; iii) \textit{Fault Localization}, which identifies the layers affected by the bit-flip attack using the Activation Shift Index (ASI); and iv) \textit{Z-Score Based Restoration}, which repairs compromised weights to recover clean functionality with minimal degradation.

\subsubsection{\textbf{AI Performance Counters (APCs)}}
We define a behavioral fingerprint (Fig.~\ref{LWM}) for each DNN layer using a set of statistical metrics derived from its activation outputs.
As shown in Fig.~\ref{fig:2}, the APC integrated with the accelerator collects key inference metrics (Table~\ref{tab:layer_metrics}). The APC block consists of a counter array, an FSM controller, and a memory block, with configuration hardware that enables dynamic counter selection for workload specific monitoring. The neuroscope, embedded within the APC, records activity per layer (e.g., sparsity, entropy, activations) in a manner conceptually similar to PyTorch hooks \cite{taylor2023extracting}, but without interfering with execution. The FSM controller enables runtime logging and adversarial detection via \textbf{PULSE} without external hooks.

The extracted metrics are summarized in Table~\ref{tab:layer_metrics}.
By tracking computational efficiency, activation sparsity, and distributional statistics, the APC enables comprehensive characterization of neural network behavior. The APC captures runtime activation statistics from each layer $\ell$ of the AI core, which are processed by the \textbf{PULSE} module of \BAREAI.

Let $\mathbf{A}_\ell \in \mathbb{R}^{B \times d}$ denote the activation tensor for a batch of $B$ inputs with $d$ activations per input.
The APC feature vector $\mathbf{v}_\ell \in \mathbb{R}^k$ comprises first- and higher-order statistics:
\[
\mathbf{v}_\ell =
\big[ \mu_\ell, \sigma^2_\ell, S_\ell, H_\ell, E_\ell, \text{skew}_\ell, K_\ell, C_\ell, F_\ell \big],
\]
where
\begin{itemize}
    \item $\mu_\ell = \tfrac{1}{Bd} \sum_{i,j} A_\ell^{(i,j)}$, \quad
          $\sigma^2_\ell = \tfrac{1}{Bd} \sum_{i,j} (A_\ell^{(i,j)} - \mu_\ell)^2$
    \item $S_\ell = \tfrac{1}{Bd} \sum_{i,j} \mathbf{1}(|A_\ell^{(i,j)}| < \epsilon)$ \hfill (sparsity)
    \item $H_\ell = - \sum_k p_k \log p_k$ \hfill (entropy)
    \item $E_\ell = \sum_{i,j} (A_\ell^{(i,j)})^2$ \hfill (activation energy)
    \item $K_\ell = \tfrac{\mathbb{E}[(A_\ell - \mu_\ell)^4]}{\sigma_\ell^4}$ \hfill (kurtosis)
    \item $C_\ell = 1 - \cos(\mathbf{A}_\ell^{\text{clean}}, \mathbf{A}_\ell^{\text{test}})$ \hfill (cosine drift)
    \item $F_\ell$ : FFT-based spectral shift in activations
\end{itemize}
Here $\mathbf{A}_\ell^{\text{clean}}$ refers to the clean reference activation profile recorded during offline profiling and stored on-chip; the clean model is never executed at runtime.
The model fingerprint is formed by all $L$ layer-wise vectors:
\[
\mathbf{V} = [\mathbf{v}_1, \ldots, \mathbf{v}_L] \in \mathbb{R}^{L \times k}.
\]

\subsubsection{\textbf{ML based Analysis Engine (PULSE)}}
In \BAREAI, the data collected by the APC is analyzed by a lightweight ML engine, \textbf{PULSE}. It continuously monitors workloads on the AI accelerator, detects unauthorized or adversarial behavior, and provides real time feedback during inference. Beyond identifying malicious activity, it alerts the user, enabling the AI IP owner to restrict model usage or specific inference capabilities. The system is model agnostic and requires no changes to the model architecture or training pipeline.

The PULSE detection function operates in two forms that we distinguish explicitly. For offline training and for all detection accuracy results reported in Section IV, PULSE is an ensemble of Random Forest, XGBoost, and LSTM classifiers trained on labeled APC feature vectors; the LSTM component captures temporal and inter-layer dynamics. For on-chip deployment and for the hardware overhead characterization in Section IV-D, the learned decision function is realized as a compact three layer neural engine with a 16-8-4 neuron configuration that operates on the selected APC features. This compact engine is the unit synthesized in the 28nm technology node. The two stage design keeps training expressive while keeping the deployed logic small enough for negligible silicon cost.

\subsubsection{\textbf{Threat Model and Security of the Monitoring Infrastructure}}
\label{sec:monitor_security}
\noindent\textbf{Attacker capabilities.}
Building on the attack surface defined in Section~\ref{sec:threat_model}, we consider an adversary who (i) can mount disturbance based fault injection such as Rowhammer~\cite{gruss2023rowhammer} and RowPress~\cite{luo2023rowpress} against off-chip DRAM, including from co-located software in virtualized settings; (ii) has white box knowledge of the victim model and can use gradient guided bit search~\cite{rakin2019bitflip,rakin2020tbfa} to select target bits; (iii) can supply arbitrary inference inputs; and (iv) may possess full knowledge of the \BAREAI~design, including which APC statistics are monitored. The attacker cannot physically probe or modify the fabricated die, cannot execute code within the trusted firmware layer that configures the monitor, and cannot directly read or write on-die monitor state. We discuss relaxations of these assumptions at the end of this section.

\noindent\textbf{Why the monitor does not share the attack surface of the weights.}
A natural question is the following: if an adversary can flip bits in the model weights, why can the same attack not corrupt the APC counters or the PULSE parameters, thereby blinding the defense? The answer lies in a fundamental asymmetry of memory technology and capacity, not in an appeal to trust. Rowhammer and its variants are disturbance phenomena specific to high density DRAM cells, in which repeated row activations induce charge leakage in physically adjacent rows \cite{kim2014flipping, mutlu2020rowhammer}. Model weights must reside in off-chip DRAM because of their size, tens to hundreds of megabytes, which places them directly within this attack surface. In contrast, the entire \BAREAI~state, comprising APC counter registers, FSM state, feature buffers, and the deployed PULSE parameters, occupies only a few kilobytes and is implemented exclusively in on-die SRAM and standard cell registers within the accelerator. On-die SRAM and flip-flops are not susceptible to Rowhammer class disturbance errors: they use a different bit cell technology, are not row activation addressable by user software, and are not mapped into any attacker accessible address space. The very property that makes the weights attackable, bulk storage in commodity DRAM, does not apply to the monitor.

\noindent\textbf{Low cost integrity hardening of the monitor.}
The small footprint of the monitor enables a complementary argument: protections that are prohibitively expensive for the model itself are essentially free for the monitor. Based on the synthesis results in Table~\ref{tab:probiotic_vs_at}, the deployed PULSE engine occupies 0.043 mm\textsuperscript{2}; triplicating it with majority voting would add approximately 0.086 mm\textsuperscript{2}, which is below 4\% of even the smallest accelerator configuration considered, and parity protection of the APC counter array and feature buffers adds a single bit per word of a few kilobyte memory. Beyond parity, the monitor state can itself be protected with error correcting codes at negligible cost: a standard (72,64) SECDED code on the PULSE parameter memory adds 12.5\% redundancy on a store of a few kilobytes, on the order of hundreds of bytes, and even multi-bit BCH codes correcting several flips per word remain trivial at this scale. We emphasize that this is fully consistent with our critique of ECC based weight protection in Section~\ref{sec:discussion}: as Eq.~\ref{eq:bch-overhead} shows, ECC cost scales with both the size of the protected data and the correction strength $t$, and is therefore prohibitive for tens of millions of weight bits yet insignificant for a kilobyte scale monitor. The same scaling law that rules out ECC for the model makes it an ideal armor for the monitor. By comparison, applying modular redundancy to tens of millions of weight parameters is infeasible by orders of magnitude. This asymmetry, a minimal trusted computing base that is cheap to harden monitoring a large asset that is expensive to harden, is the standard rationale for hardware roots of trust, and \BAREAI~follows the same principle. In addition, the deployed PULSE parameters can be periodically validated against a reference digest held in on-die non-volatile or fuse based storage; any mismatch triggers an alert and quarantines the inference output. APC configuration registers are writable only through the trusted firmware path established at secure boot, so a user space or guest level attacker cannot lower thresholds or disable counters.

\noindent\textbf{Side channel considerations.}
The monitoring infrastructure raises two side channel questions. First, the monitor could itself become a leakage source: APC traces summarize per-layer activation statistics and could, if exposed, leak information about the model architecture or the inputs being processed. \BAREAI~therefore never exports raw traces off-die. APC outputs are consumed directly by PULSE on the same silicon, and only the binary detection verdict, optionally with the flagged layer index, crosses the trust boundary. Raw counter access is restricted to the trusted firmware interface, and readouts can be quantized and temporally aggregated so that fine grained per-inference values are not observable even to privileged diagnostic software. We note that the switching activity of the counter logic is small relative to the MAC array and memory subsystem, limiting its visibility in the power signature; a formal power and electromagnetic side channel evaluation of the monitor is left to future work. Second, an attacker who learns the monitored statistics or thresholds may attempt side channel assisted evasion, crafting flips that remain below the detection boundary. Our adaptive attack evaluation (Table~\ref{tab:multi_bit_ttack}) models a strong version of this adversary: gradient guided attacks that jointly optimize bit positions and inputs to suppress APC deviations, with knowledge of the feature set, are detected with at least 93.8\% accuracy. Two design properties limit evasion further. The detection boundary is not a single public constant: the adaptive threshold is calibrated per device and per model from local clean run statistics, so thresholds differ across deployments. Randomized temporal sampling of APC readouts additionally denies the attacker a deterministic view of which inference windows are being scored.

\noindent\textbf{Limitations under weaker threat models.}
For transparency, we state the boundaries of our guarantees. If the adversary compromises the trusted firmware layer itself, for example through an exploited secure boot chain, the APC configuration interface is exposed and \BAREAI~degrades to the protection level of the underlying platform; defending the firmware root of trust is an orthogonal problem addressed by measured and verified boot. A hardware Trojan inserted at design or fabrication time could, by construction, bypass any on-die monitor including ours; pre-silicon verification and post-silicon Trojan detection are complementary lines of defense. Invasive physical attacks such as microprobing are outside our model, as they defeat all non tamper resistant hardware equally. Within the stated model, a software level or disturbance based fault injection adversary with white box model knowledge, the monitor attack surface is physically disjoint from that of the weights, and its integrity can be maintained at negligible cost.

\subsection{\textbf{Bit-flip Attack Detection \& Mitigation}}
The AI accelerator supports both training and inference. During inference, the APC block logs performance metrics (Table~\ref{tab:layer_metrics}) in dedicated memory and forwards them to \textbf{PULSE} for activity monitoring. On detecting anomalies, \textbf{PULSE} flags the incident, can halt the affected IP block, and reports to the oversight authority. Adversarial detection in \BAREAI~requires selecting counters during an offline phase using clean and adversarial samples. Feature selection identifies the most informative APCs, which remain fixed during inference; in certain applications, data independent APC choices may also be viable. The complete runtime defense workflow of \BAREAI~is described in Algorithm~\ref{alg:bareai}. It defends against BFAs by leveraging activation based APCs through a two phase process (offline and online) that is model agnostic and hardware efficient.

\BAREAI~incorporates a two stage normalization and adaptive thresholding mechanism. Z-score normalization standardizes each APC metric per layer during offline profiling using the mean and variance from clean executions, centering benign variation and scaling dispersion to ensure well conditioned features for the PULSE detector.
The adaptive threshold $\tau_{\text{adaptive}}$ defines a dynamic anomaly boundary using clean signature statistics, with $\tau_{\text{adaptive}} = \mu_c + \lambda\,\sigma_c$,
where $\mu_c$ and $\sigma_c$ denote the mean and standard deviation of clean signatures, and $\lambda$ (empirically set to 2) controls sensitivity to benign drift.
This enables PULSE to adapt detection confidence to model and dataset specific variability. Additionally, PULSE stabilizes detection and reduces false positives by aggregating Random Forest, XGBoost, and LSTM outputs through weighted voting.

\begin{algorithm}[!ht]
\small
\caption{\textsc{Bit-Flip Detection and Mitigation (\BAREAI)}}
\label{alg:bareai}
\DontPrintSemicolon
\LinesNotNumbered
\SetKwInOut{Input}{Input}
\SetKwInOut{Output}{Output}

\Input{
  Clean reference model $M_{\text{clean}}$ (offline only); \\
  Test model $M_{\text{test}}$ (potentially attacked); \\
  Dataset $\mathcal{D}=\{(x_i,y_i)\}_{i=1}^N$
}
\Output{
  Detector $\mathcal{D}_{\text{PULSE}}$; Mitigated model $M_{\text{fixed}}$
}

\tcc{Step 1 (offline): APC Feature Extraction and Profiling}
\ForEach{$(x_i,y_i)\in\mathcal{D}$}{
  \ForEach{layer $\ell$ of $M_{\text{clean}}$ and $M_{\text{test}}$}{
    Extract activations $a^{(\ell)}$ \\
    Compute APC feature vector $\mathbf{v}_\ell$ ($\mu$, $\sigma^2$, $H$, $E$, etc.) \\
    Store clean reference statistics $\mu^{\text{clean}}_\ell$, $\sigma^{\text{clean}}_\ell$, $\mathbf{v}^{\text{clean}}_\ell$ \\
  }
}
Store clean vs. attacked features $\mathcal{F}_{\text{clean}},\mathcal{F}_{\text{adv}}$ with labels.

\tcc{Step 2 (offline): PULSE Training}
Train detector $\mathcal{D}_{\text{PULSE}}$ on $(\mathcal{F},y)$ pairs using the RF, XGBoost, and LSTM ensemble; distill to the compact on-chip engine.

\tcc{Step 3 (online): Attack Detection via PULSE}
\ForEach{test input $x$}{
  Derive $\mathcal{F}_x$ from runtime APCs of $M_{\text{test}}$, using stored clean reference statistics for drift features \\
  Predict $\hat{y}=\mathcal{D}_{\text{PULSE}}(\mathcal{F}_x)$ \\
  \If{$\hat{y}=0$}{Model benign; stop.}
}

\tcc{Step 4 (online): Localization via ASI}
\ForEach{layer $\ell$}{
  $\text{ASI}_\ell = \dfrac{1}{k}\sum_{j=1}^{k}\left|\dfrac{v^{\text{test}}_{\ell,j}-v^{\text{clean}}_{\ell,j}}{v^{\text{clean}}_{\ell,j}+\delta}\right|$ \\
  $\Delta_\ell = \text{ASI}_\ell - \text{ASI}^{\text{clean}}_\ell$ \\
}
Flag faulty layers $\mathcal{L}_{\text{fault}} = \{\ell \mid \Delta_\ell > \mu_\Delta + 2\sigma_\Delta\}$

\tcc{Step 5 (online): z-score repair}
\ForEach{$\ell\in\mathcal{L}_{\text{fault}}$}{
  For each weight $w \in W_\ell$: \\
  Compute $Z = \dfrac{w - \mu^{\text{clean}}_\ell}{\sigma^{\text{clean}}_\ell}$ \\
  \eIf{$|Z|>\tau_z$}{reset $w \leftarrow \mu^{\text{clean}}_\ell$}{keep $w$ unchanged.}
}
Obtain repaired model $M_{\text{fixed}}$.

\Return{$\mathcal{D}_{\text{PULSE}}, M_{\text{fixed}}$}
\end{algorithm}

\subsubsection{\textbf{APC Feature Extraction}}
During the offline phase [Step 1], for each input sample $(x_i, y_i)$ in the dataset $\mathcal{D}$, both the clean model $M_{\text{clean}}$ and an adversarially attacked model $M_{\text{adv}}$ are executed.
At every layer $\ell$, we extract the activation tensor $a^{(\ell)}$ and compute the APC features: first order statistics (mean $\mu$, variance $\sigma^2$), sparsity (fraction of near zero activations), entropy $H(a)$, energy $E(a) = \sum_{i,j} (a^{(\ell)}_{i,j})^2$, and higher order moments (skewness and kurtosis). These features form a compact behavioral signature per layer. The per-layer clean reference statistics ($\mu^{\text{clean}}_\ell$, $\sigma^{\text{clean}}_\ell$, and the reference feature vector $\mathbf{v}^{\text{clean}}_\ell$) are stored for use at runtime; this requires only a few scalars per layer and does not involve retaining a duplicate model.
The APC vectors are labeled with $y = 0$ (clean) or $y = 1$ (compromised) and stored as $\mathcal{F}_{\text{clean}}, \mathcal{F}_{\text{adv}}$, forming the training set for the classifier.

\subsubsection{\textbf{PULSE Training Process}}
\begin{figure*}[h]
    \captionsetup{labelfont=bf}
    \centering
    \includegraphics[width=0.7\textwidth]{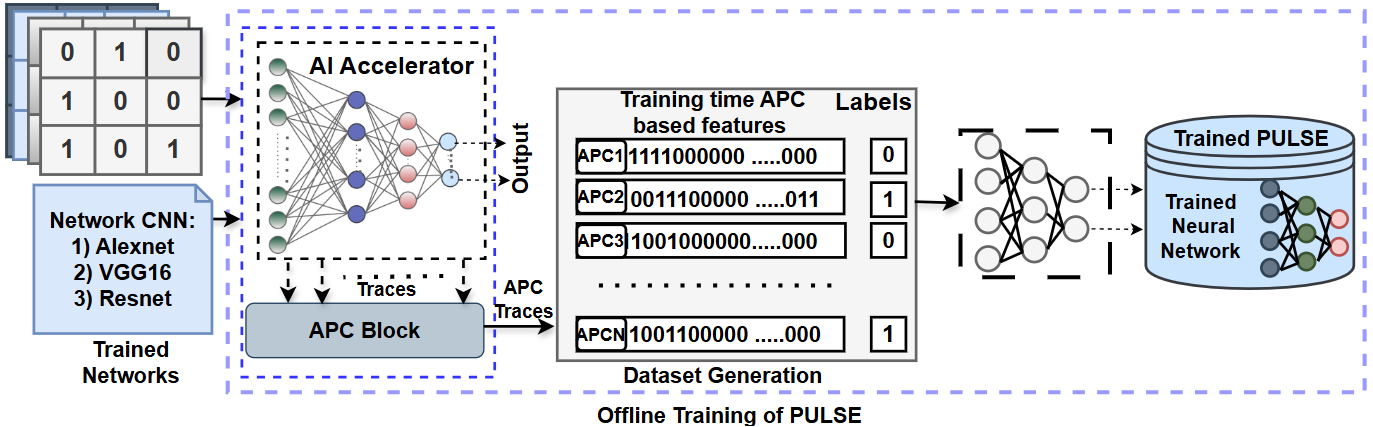}
    \caption{Diagram of the \textbf{PULSE} training workflow. During inference, APCs capture runtime traces of the AI accelerator (e.g., sparsity, entropy, gradient norms) for both clean and adversarial data. Important features from these traces are used to train \textbf{PULSE}, a detector that distinguishes clean and tampered models for real-time bit-flip attack detection.}
    \label{fig:2}
\end{figure*}

\textbf{PULSE} training is shown in Fig.~\ref{fig:2}. The AI accelerator runs inference on both clean and adversarial data, while the APC captures operational traces under each condition. These traces form the training dataset for PULSE, using standard vision benchmarks and CNN models such as ResNet, VGG, and AlexNet. APC derived features enable PULSE to learn the distinctions between normal and adversarial behavior. Using the labeled feature vectors $(\mathcal{F}, y)$, we train the detector as an ensemble of Random Forest, XGBoost, and LSTM classifiers [Step 2]; the LSTM captures inter-layer temporal dynamics. The learned decision function is then distilled into the compact on-chip engine described in Section III-A.2 for deployment. Once deployed, PULSE monitors APC traces at runtime and activates safeguards upon detecting adversarial patterns.

\subsubsection{\textbf{BFA Attack Detection via PULSE}}
During inference, the AI accelerator processes unlabeled inputs while the APCs record the metrics of Table~\ref{tab:layer_metrics}. PULSE analyzes these traces and compares deviations from profiled normal behavior to detect adversarial patterns in real time. Upon detection, PULSE initiates security measures such as alerts or halting execution.

Operationally [Step 3], for each inference time sample $x_i$ on the protected model $M_{\text{test}}$: i) we run $x_i$ on $M_{\text{test}}$ and extract runtime APC values; ii) we compute deviation features (e.g., $\Delta\mu$, $\Delta$Sparsity, ASI, cosine drift) against the stored clean reference statistics from offline profiling; iii) the APC feature vector $\mathcal{F}_{x_i}$ is formed; iv) PULSE predicts $\hat{y}_i = \mathcal{D}_{\text{PULSE}}(\mathcal{F}_{x_i})$.
If $\hat{y}_i = 1$, the system flags $M_{\text{test}}$ as compromised and optionally triggers repair.

We emphasize that \BAREAI~runs a single model during inference. The clean model is executed only offline to generate the reference statistics and the PULSE training data. At runtime there is no duplicate model; drift features such as cosine drift are computed against the stored reference profile, not against a live clean execution. PULSE is a lightweight auxiliary classifier that analyzes real time APC features without executing or replicating the DNN.

\subsubsection{\textbf{Attack Layer Localization}}
To localize the layers affected by bit flips, we introduce the \textbf{Activation Shift Index} (ASI). For a layer $\ell$, the ASI quantifies the normalized deviation of the runtime APC feature vector from the stored clean reference profile:
\[
\text{ASI}_\ell = \frac{1}{k} \sum_{j=1}^{k} \left| \frac{v_{\ell,j}^{\text{test}} - v_{\ell,j}^{\text{clean}}}{v_{\ell,j}^{\text{clean}} + \delta} \right|
\]
where $k$ is the number of APC features and $\delta$ a small constant for numerical stability. A higher ASI indicates a greater behavioral anomaly.

This stage pinpoints which layers are responsible for the anomalous behavior [Step 4]. We compute the per-layer delta $\Delta_\ell = \text{ASI}_\ell - \text{ASI}_\ell^{\text{clean}}$, where $\text{ASI}_\ell^{\text{clean}}$ is the baseline value of the index observed during clean profiling; the delta captures shifts beyond natural variance. A statistical threshold derived from the distribution of $\Delta_\ell$ identifies layers with excessive shifts as
$\mathcal{L}_{\text{fault}} = \left\{ \ell \,\middle|\, \Delta_\ell > \mu_\Delta + 2\sigma_\Delta \right\}$.
Only these localized layers are candidates for mitigation.

\subsubsection{\textbf{Attack Mitigation}}
After fault localization, anomalous weights in the flagged layers are repaired using a z-score filter against the clean layer statistics stored during offline profiling. For a flagged layer $\ell$ with weights $\theta_\ell$, the z-score of a weight $w_i \in \theta_\ell$ is
\[
z(w_i) = \frac{w_i - \mu^{\text{clean}}_\ell}{\sigma^{\text{clean}}_\ell}.
\]
Weights whose z-score magnitude exceeds a threshold $\tau_z$ are considered anomalous and reset to the clean layer mean:
\[
w_i^{\text{fixed}} =
\begin{cases}
\mu^{\text{clean}}_\ell & \text{if } |z(w_i)| > \tau_z, \\
w_i & \text{otherwise}.
\end{cases}
\]
This statistical filtering removes the extreme outliers introduced by bit flips, most notably exponent and sign flips that push weights far outside the clean distribution, while leaving benign parameters untouched. The scheme requires storing only two scalars per layer ($\mu^{\text{clean}}_\ell$ and $\sigma^{\text{clean}}_\ell$); no golden copy of the weights is retained. We note the inherent trade-off of mean substitution: a repaired weight recovers statistical regularity but not its exact original value, which is why recovery is strongest when the flipped weights are extreme outliers (Section IV-C).

Overall, Algorithm~\ref{alg:bareai} unifies detection, localization, and mitigation into an efficient runtime system. By combining statistical profiling (APCs, ASI), the PULSE detection engine, and explainable recovery (z-score filtering), it provides a practical defense against bit level tampering in real world AI deployments.

\section{\textbf{Experimental Results}}
We evaluate the effectiveness of the proposed approach under a wide range of adversarial bit-flip attacks.
\vspace{-5mm}
\subsection{\textbf{Experimental Setup}}
To evaluate \BAREAI, we used a diverse set of neural network architectures and datasets widely adopted in adversarial robustness research. The CNN models include AlexNet, VGG-16, ResNet-18, and ResNet-50, covering lightweight and deep architectures. We further extended the evaluation to ViT-B and MobileNetV2 to capture performance across transformer and efficient convolutional architectures.
The vision models were benchmarked on standard datasets such as ImageNet-1K \cite{imagenet}. We also assess \BAREAI~on Large Language Models using LLaMA-3.2-1B~\cite{llama} on the Massive Multitask Language Understanding (MMLU)~\cite{mmlu} benchmark, along with additional language model evaluations reported in the corresponding tables. These workloads collectively stress test the framework across heterogeneous data modalities.
To simulate adversarial threats, we implemented state of the art BFAs~\cite{rakin2019bitflip}, including random bit flips and Progressive Bit Search (PBS)~\cite{rakin2020tbfa}. These attacks represent both random and highly targeted strategies, enabling a comprehensive robustness evaluation.

The detection and mitigation pipeline was evaluated on an NVIDIA RTX A6000 GPU platform. This platform was chosen for algorithmic validation of runtime detection accuracy across diverse architectures and for fair comparison with prior bit-flip defenses, which also employ GPU based evaluations; in this setting, APC feature extraction is emulated in software with hooks that mirror the hardware counter semantics. The hardware overhead numbers reported in Section IV-D are derived exclusively from 28nm synthesis of the \BAREAI~subsystem and do not rely on GPU profiling. We state this division explicitly: detection and recovery accuracies come from the GPU based software evaluation of the PULSE ensemble, while area, energy, and timing overheads come from synthesis of the deployed monitor.

To quantify detection and mitigation performance, we measured detection accuracy, false positives, false negatives, and classification recovery under attack. Hardware level metrics of \BAREAI, including energy, area, and timing, were obtained by synthesizing the APC and PULSE subsystem on a 28nm CMOS process using Synopsys Design Compiler. \BAREAI~operates alongside target DNN accelerator architectures which are not re-synthesized; instead, published Eyeriss silicon measurements serve as the accelerator baseline (Section IV-D). For adversarial detection, \BAREAI~extracts runtime metrics (sparsity, entropy, activation energy, cosine drift, kurtosis, etc.), which are processed by the PULSE ensemble (Random Forest, XGBoost, and LSTM). For explainability and fault localization, we employ the ASI and z-score based restoration. All experiments were repeated multiple times, and reported results are averages across multiple attack seeds. For comparison, we also evaluated ECC based baselines to characterize the trade-off between detection capability and overhead.

\begin{figure}[!t]
\label{fig:apc_dist}
    \centering
    \captionsetup{labelfont=bf}
    \begin{subfigure}[t]{0.85\linewidth}
        \includegraphics[width=0.85\linewidth]{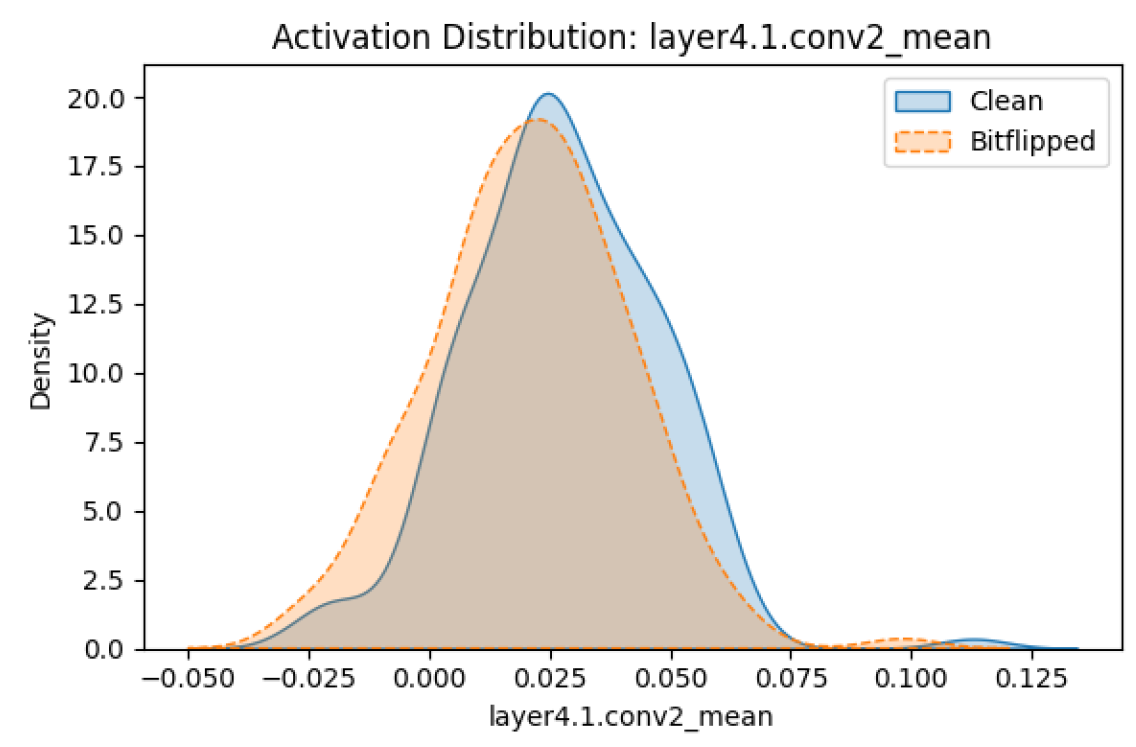}
        \caption{Activation distribution.}
        \label{fig:activation_dist}
    \end{subfigure}
    \begin{subfigure}[t]{0.85\linewidth}
        \includegraphics[width=0.85\linewidth]{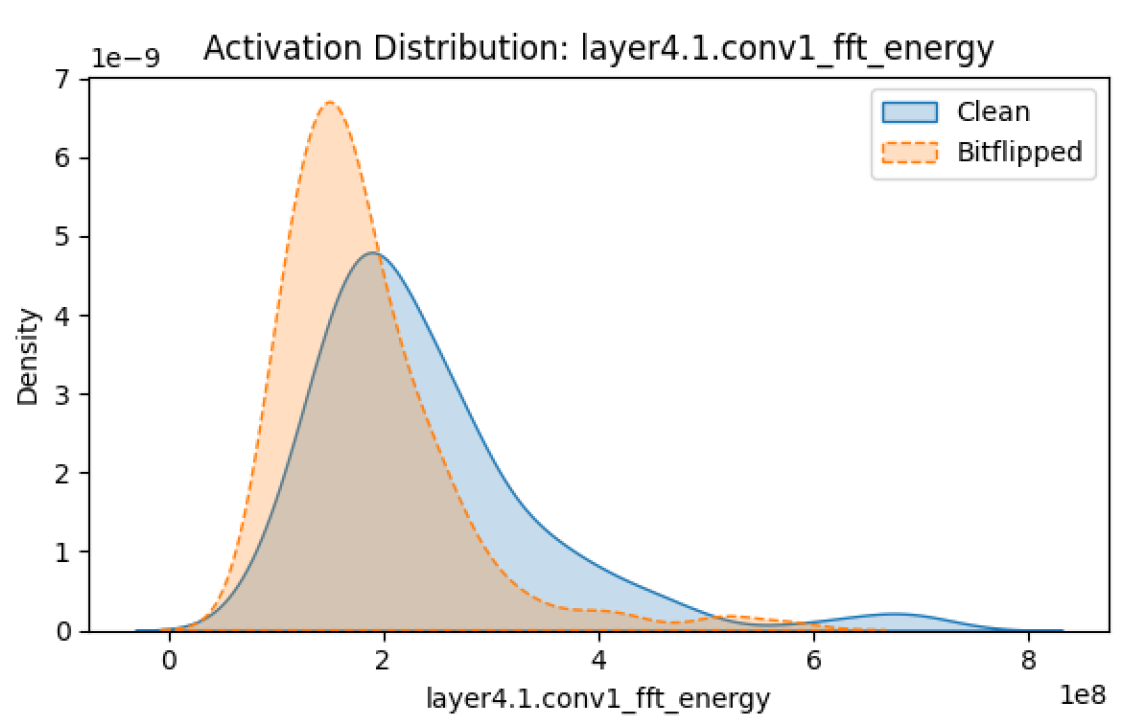}
    \caption{Activation energy distribution.}
    \label{fig:apc_trace}
    \end{subfigure}
    \caption{Difference in distribution of activations and their FFT-based energy.}
\end{figure}

\begin{figure}[!b]
    \centering
    \captionsetup{labelfont=bf}
    \includegraphics[width=0.90\linewidth]{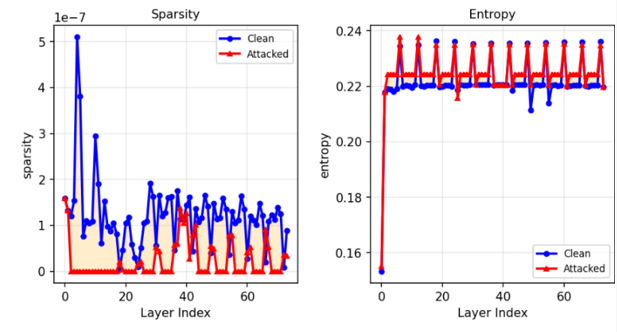}
    \caption{Layer-wise sparsity (left) and entropy (right) comparison for clean vs. attacked models, showing suppressed sparsity under attack.}
    \label{fig:apc_sparsity}
\end{figure}

\begin{figure}[ht]
    \centering
    \captionsetup{labelfont=bf}
    \includegraphics[width=1 \linewidth]{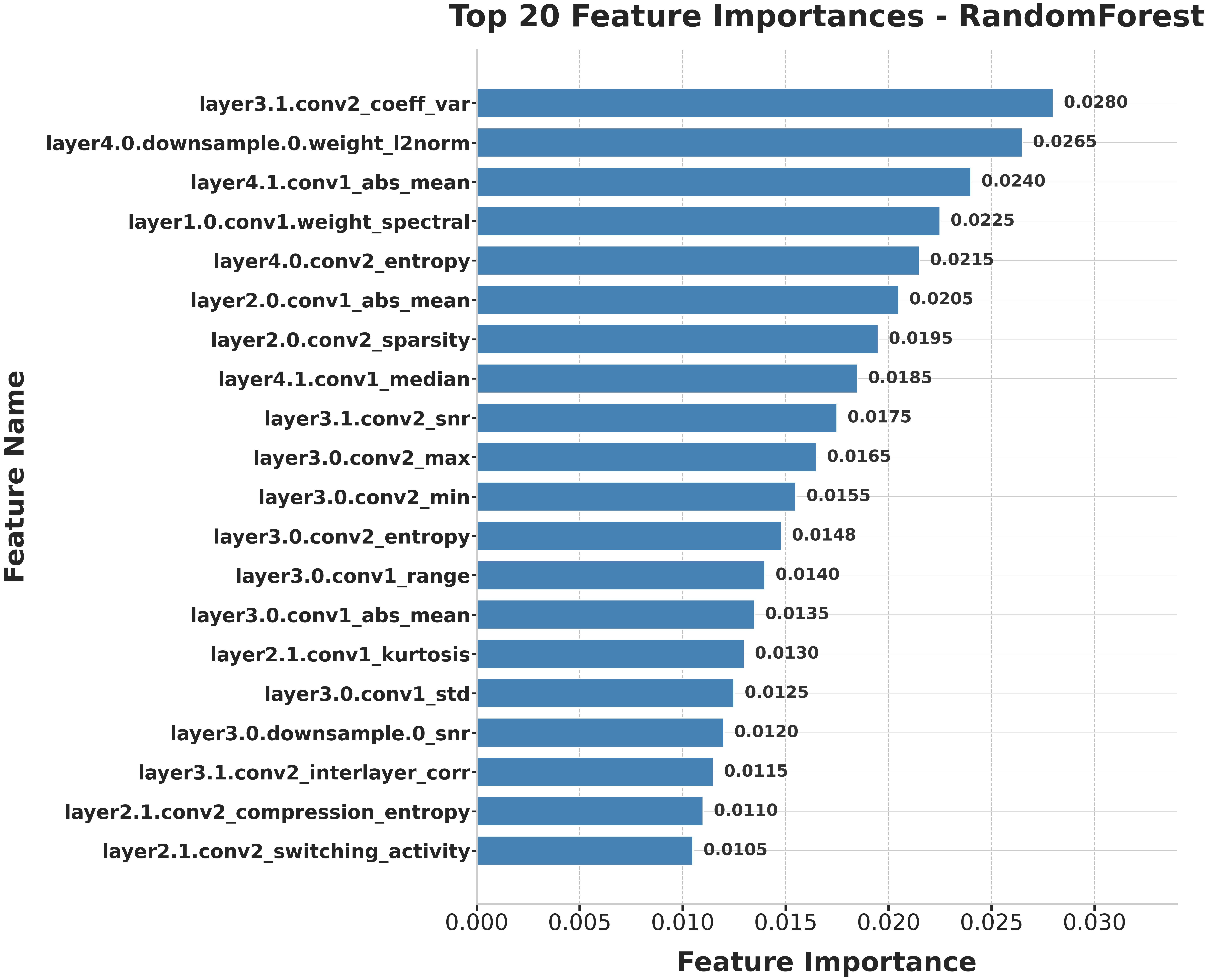}
    \caption{Feature importance ranking of APC metrics using Random Forest, highlighting sparsity, loss, and norm-based statistics as dominant contributors.}
    \label{fig:feature_importance}
\end{figure}

\vspace{-1em}
\subsection{\textbf{Choosing Critical \BAREAI~APCs for Attack Detection}}
\label{subsec:explain_features_results}
Figures~\ref{fig:activation_dist} through \ref{fig:feature_importance} illustrate how bit-flip perturbations manifest in layer activations, generating distinctive APC traces that \BAREAI~leverages for detection. Figure~\ref{fig:activation_dist} compares the activation distributions of clean and bit-flipped models. Even with limited perturbations, the attacked model exhibits subtle but consistent shifts in mean activation values, confirming that bit flips propagate as small yet measurable changes in layer statistics. These deviations, though not visually evident in the weight distributions, accumulate across layers and destabilize downstream predictions.

Figure~\ref{fig:apc_trace} further highlights the sensitivity of APC metrics by analyzing FFT based activation energy. The sparsity trace shows a clear drop in near zero activations under attack, whereas entropy remains relatively stable, suggesting that sparsity serves as a stronger discriminator than entropy for certain layers. Together, these results validate the design choice to employ multiple complementary APC features to capture diverse forms of perturbation.

Figure~\ref{fig:apc_sparsity} extends this analysis to the full depth of the network. Layer wise sparsity diverges significantly between clean and attacked models, with the attacked trace showing consistently suppressed sparsity values. In contrast, entropy remains tightly aligned between clean and attacked runs, underscoring that not all counters shift equally. Such feature diversity enables the ensemble detector \textbf{PULSE} to selectively weight the most discriminative metrics.

Finally, Figure~\ref{fig:feature_importance} reports the top 20 APC features ranked by Random Forest importance scores.
Importance values are aggregated by feature category rather than individual metrics. The 20 APC features were grouped into (i) statistical and distribution based metrics (mean, variance, sparsity, loss, activation energy, etc.) and (ii) higher order and spectral metrics (L1/L2 norm, MSE, kurtosis, entropy, cosine drift, FFT shift, etc.). The Random Forest module computes group wise importance to minimize collinearity and enable model agnostic comparison, so Figure~\ref{fig:feature_importance} shows category level significance.
Metrics such as loss, mean square error, $L_{1}/L_{2}$ norms, sparsity, and zero ratios dominate the ranking, indicating that both global performance indicators and fine grained activation statistics contribute meaningfully to attack detection. This confirms that the detection efficacy of \BAREAI~stems from heterogeneous counter values rather than any single metric.

\begin{table*}[h]
\centering
\captionsetup{labelfont=bf}
\caption{~\BAREAI~ Bit-Flip Attack Detection Accuracy for ViT-B, ResNet-50, and MobileNetV2 (ImageNet-1K)}
\label{tab:attack_types}
\renewcommand{\arraystretch}{1} 
\setlength{\tabcolsep}{12pt}      
\begin{tabular}{@{}lccc|ccc|ccc@{}}
\toprule
\textbf{Attack Type}
& \multicolumn{3}{c|}{\textbf{ViT-B}}
& \multicolumn{3}{c|}{\textbf{ResNet-50}}
& \multicolumn{3}{c}{\textbf{MobileNetV2}} \\
\cmidrule(lr){2-4} \cmidrule(lr){5-7} \cmidrule(lr){8-10}
& \textbf{Orig.} & \textbf{Attacked} & \textbf{Detect.}
& \textbf{Orig.} & \textbf{Attacked} & \textbf{Detect.}
& \textbf{Orig.} & \textbf{Attacked} & \textbf{Detect.} \\
\midrule
BFA  & 85\% & 42\% & 93\% & 80\% & 38\% & 92\% & 78\% & 35\% & 91\% \\
RBFA & 85\% & 50\% & 90\% & 80\% & 47\% & 89\% & 78\% & 45\% & 88\% \\
PBS  & 85\% & 25\% & 96\% & 80\% & 22\% & 95\% & 78\% & 20\% & 95\% \\
\midrule
\textbf{Avg.}
& 85\% & 39\% & 93\%
& 80\% & 36\% & 92\%
& 78\% & 33\% & 91\% \\
\bottomrule
\end{tabular}
\end{table*}

\vspace{-1em}
\subsection{\textbf{\BAREAI~Performance Analysis}}
In this section, we use the insights obtained from the APC metrics (Section~\ref{subsec:explain_features_results}) to systematically detect and localize BFAs in different DNN models, mitigate their effects, and assess the effectiveness of \BAREAI.

\subsubsection{\textbf{Detection Performance}}
To assess the capability of \BAREAI~in detecting bit-flip attacks, we experimentally investigate its detection performance and robustness in detail.
\vspace{1mm}

\noindent{\textbf{Detection Accuracy of \BAREAI:}}
Table~\ref{tab:attack_types} compares the impact of three representative bit-flip attack strategies, Bit-flip Attack (BFA), Random Bit-Flip Attack (RBFA), and Progressive Bit Search (PBS), on ViT-B, ResNet-50, and MobileNetV2 using ImageNet-1K. Random bit flips moderately reduce accuracy, while sensitivity guided BFA amplifies degradation. The most destructive case is PBS, which selectively flips the most sensitive weight bits based on gradient information, reducing ResNet-50 accuracy from 80\% to only 22\%. Despite such severe degradation, \BAREAI~consistently achieves high detection accuracy across all models, with average detection rates of 93\% for ViT-B, 92\% for ResNet-50, and 91\% for MobileNetV2.
These results show that \BAREAI~is model agnostic and effective across both convolutional and transformer architectures.

\begin{table}[h]
\centering
\captionsetup{labelfont=bf}
\caption{Detection Performance Across Bit Groups}
\label{tab:bit-groups}
\renewcommand{\arraystretch}{1.0}
\setlength{\tabcolsep}{3pt}
\begin{tabular}{@{}lccc>{\raggedright\arraybackslash}p{0.3\linewidth}@{}}
\toprule
\textbf{Bit Group} & \textbf{Det. Acc.} & \textbf{FP$^\ast$} & \textbf{FN$^\ast$} & \textbf{Impact / Remarks} \\
\midrule
Exponent (FP30--23) & 95--98\% & 2--3\% & 1--2\% & Very high distortion; \newline easily detectable \\
Sign Bit (FP31)     & 90--93\% & ~5\%   & 5--7\% & High; flips activation \newline direction, alters \newline mean/kurtosis \\
Mantissa (FP22--0)  & 70--75\% & 10--15\% & 15--20\% & Low--moderate; small \newline shifts, patterns \newline look normal \\
\bottomrule
\end{tabular}

\vspace{4pt}
\parbox{0.5\textwidth}{\footnotesize\centering $^\ast$FP = False Positives,\; FN = False Negatives}
\end{table}
\vspace{1mm}

\noindent{\textbf{Detecting BFAs on different bit groups:}}
Although detection rates are generally high, Table~\ref{tab:bit-groups} shows clear differences between floating point bit groups. Exponent flips (FP30 to FP23) cause large distortions and are detected with 95 to 98\% accuracy and only 1 to 2\% false negatives. Sign bit flips (FP31) are moderately harder, achieving 90 to 93\% accuracy with up to 7\% false negatives, due to changes in activation direction and statistics. Mantissa flips (FP22 to FP0) are the most subtle, reducing detection accuracy to 70 to 75\% and increasing false negatives to 15 to 20\%. These results characterize both the strength of \BAREAI~against high impact perturbations and its current limits against low magnitude mantissa flips, which by construction cause small behavioral deviation and correspondingly small model damage.
\vspace{1mm}

\noindent{\textbf{Cross Attack Detection Performance:}}
Our APC feature selection is attack agnostic, relying on fundamental activation level statistics (e.g., entropy drift, sparsity ratio, cosine similarity shift) that capture deviations from weight perturbations independent of the attack method. These intrinsic properties generalize to unseen attacks. As shown in Table~\ref{tab:cross_attack}, training PULSE on one attack (Progressive Bit Search) and testing on another (Random Bit-Flip) results in only a 1.7\% accuracy drop, demonstrating strong cross attack robustness. Moreover, PULSE supports incremental updates with new APC traces without retraining the base model, enabling continual adaptation to emerging attack variants.

\begin{table}[h]
\tiny
\centering
\captionsetup{labelfont=bf}
\caption{Cross Attack Detection Performance of PULSE}
\label{tab:cross_attack}
\resizebox{\columnwidth}{!}{%
\begin{tabular}{lcc}
\hline
\multicolumn{1}{c}{\textbf{\begin{tabular}[c]{@{}c@{}}Training Attack \\ Type\end{tabular}}} &
  \textbf{\begin{tabular}[c]{@{}c@{}}Testing Attack \\ Type\end{tabular}} &
  \textbf{\begin{tabular}[c]{@{}c@{}}Detection \\ Accuracy (\%)\end{tabular}} \\ \hline
Progressive Bit Search (PBS) & Same (PBS)            & 95.5           \\
Progressive Bit Search (PBS) & Random Bit-Flip (RBF) & 93.8           \\ \hline
\multicolumn{2}{r}{\textbf{Accuracy Degradation}}    & \textbf{1.7\%} \\ \hline
\end{tabular}%
}
\end{table}

\noindent{\textbf{Detection Performance for Cross Domain Attack:}}
PULSE generalizes beyond dataset specific biases by using architecture agnostic APC features and ensemble learning that capture intrinsic activation dynamics rather than pixel semantics. We evaluate this by training PULSE on one dataset and model pair and testing directly on different datasets without retraining. Trained across diverse architectures (ResNet, MobileNet, ViT) and datasets (CIFAR-10, ImageNet, Tiny-ImageNet), PULSE learns a high order decision boundary in the APC space using the RF, XGB, and LSTM ensemble. In cross domain tests (e.g., ImageNet to CIFAR-100 and SVHN), average accuracy reaches 94.8\% (Table~\ref{tab:my-table-12}). Cross architecture transfer between CNNs and ViTs incurs under a 3\% accuracy drop, confirming robust generalization.

\begin{table}[h]
\fontsize{40pt}{40pt}\selectfont
\centering
\captionsetup{labelfont=bf}
\caption{Cross-domain and Cross-architecture Generalization of PULSE}
\renewcommand{\arraystretch}{1.2}
\label{tab:my-table-12}
\resizebox{\columnwidth}{!}{%
\begin{tabular}{llll}
\hline
\textbf{Training Domain} & \textbf{Testing Domain} & \textbf{Detection Acc. (\%)} & \textbf{FNR (\%)} \\ \hline
ImageNet / ResNet-50   & CIFAR-100 / ResNet-18       & 95.2          & 4.1          \\
ImageNet / ResNet-50   & SVHN / ResNet-18            & 94.3          & 4.8          \\
CIFAR-10 / MobileNetV2 & Tiny-ImageNet / MobileNetV2 & 95.7          & 3.9          \\
ImageNet / ViT-B       & CIFAR-10 / ResNet-18        & 93.8          & 5.4          \\
CIFAR-100 / ResNet-18  & ImageNet / ViT-B            & 94.9          & 4.6          \\ \hline
\textbf{Average}       &                             & \textbf{94.8} & \textbf{4.6} \\ \hline
\end{tabular}%
}
\end{table}

\noindent{\textbf{Detection Performance for Small Random Bit-Flips:}}
Detecting subtle perturbations is critical for runtime robustness. To evaluate \textbf{PULSE} under minimal corruption, we injected random low rate bit-flip faults ($N_{\text{flip}}\!\in\!\{1,2,4,8\}$, below 0.001\% of total weights) into ResNet-18 and ViT-B across 200 randomized trials per configuration. Table~\ref{tab:lowrate} shows that detection accuracy exceeds 91\% for a single flip and rises to 97.4\% for eight flips, with false alarm rates below 4\%. We observe that (i) spectral and entropy features are highly sensitive, providing early fault cues; and (ii) mean and sparsity metrics tolerate small noise, showing limited sensitivity to single flips. Overall, PULSE reliably detects even minute perturbations without excess false positives; ensemble fusion mitigates metric level masking effects.

\begin{table}[!ht]
\large
\centering
\captionsetup{labelfont=bf}
\caption{Detection Accuracy for Small Random Bit-Flips}
\renewcommand{\arraystretch}{1.5}
\setlength{\tabcolsep}{1pt}
\label{tab:lowrate}
\resizebox{\columnwidth}{!}{%
\begin{tabular}{llll}
\hline
\textbf{No. of Flips ($N_{\text{flip}}$)} & \textbf{Detection Acc. (\%)} & \textbf{False Alarm (\%)} & \textbf{Dominant Features} \\ \hline
1 & 91.3 & 4.2 & Entropy, Spectral Shift \\
2 & 93.8 & 3.8 & Entropy, Kurtosis \\
4 & 96.1 & 3.2 & Cosine Drift, Variance \\
8 & 97.4 & 2.9 & Combined (All Domains) \\ \hline
\end{tabular}
}
\vspace{1mm}
\end{table}

\noindent{\textbf{Performance under Distributed Multi-bit-flip Attacks:}}
Advanced multi-bit, distributed, or adaptive faults can obscure layer wise APC deviations. To assess the resilience of \textbf{PULSE}, we evaluated two challenging scenarios: (i) spatially distributed bit flips across 3 to 6 layers, and (ii) adaptive gradient guided attacks that jointly optimize bit positions and inputs to minimize APC anomalies. In ResNet-18, MobileNetV2, and ViT-B, 4 to 12 bit flips were injected into convolutional and fully connected layers. Despite these correlated perturbations, PULSE maintained a detection accuracy of 95.1 to 97.6\% with false negatives below 4.5\%, leveraging LSTM based temporal aggregation and cross layer covariance checks to capture subtle and persistent deviations.

\noindent{\textbf{Performance against Adaptive bit-flip Attacks:}}
For adaptive attacks (e.g., DeepSearch-BFA and Adaptive BitFlip) that jointly optimize bit positions and gradients to suppress APC shifts, \textbf{PULSE} maintains detection accuracy of at least 93.8\% with false negative rates below 5.8\%. Its spatial and temporal ensemble reasoning and adaptive thresholding capture structured drifts that deviate from natural activation dynamics, limiting evasion. Even when perturbations are spatially distributed or optimized to mimic benign behavior, residual inconsistencies in temporal APC evolution expose the attack. As summarized in Table~\ref{tab:multi_bit_ttack}, \textbf{PULSE} consistently achieves above 93\% accuracy with below 5.8\% FNR and about 6 cycle decision latency, providing robust protection against coordinated and optimization driven multi-bit faults.
\begin{table}[h]
\Large
\centering
\captionsetup{labelfont=bf}
\caption{Detection Performance under Multi-bit and Distributed Bit-flip Attacks}
\renewcommand{\arraystretch}{1.3}
\label{tab:multi_bit_ttack}
\resizebox{\columnwidth}{!}{%
\begin{tabular}{lccc}
\hline
\textbf{Attack Type / Model} & \textbf{Detection Acc. (\%)} & \textbf{FNR (\%)} & \textbf{Latency (cycles)} \\ \hline
Distributed (6 bits, ResNet-18)    & 97.2          & 3.8          & 5.6          \\
Distributed (12 bits, MobileNetV2) & 95.8          & 4.5          & 6.1          \\
Adaptive Gradient (ResNet-18)      & 94.3          & 5.1          & 5.9          \\
Adaptive Gradient (ViT-B)          & 93.8          & 5.8          & 6.4          \\ \hline
\textbf{Average}                   & \textbf{95.3} & \textbf{4.8} & \textbf{6.0} \\ \hline
\end{tabular}%
}
\end{table}

\subsubsection{\textbf{Attack Layer Localization}}
Table~\ref{tab:anomaly_scores} highlights the efficacy of \BAREAI~in localizing the layers most impacted by a bit-flip attack using the Activation Shift Index. For ResNet-18, the highest anomaly appears at the first convolution layer, with additional deviations in mid residual blocks; this correlates with a severe accuracy drop of 66.12\%, showing that CNNs are highly sensitive to early layer perturbations. In contrast, a pretrained BERT-Large model on the SQuAD question answering dataset records its maximum anomaly score of 38.191 at \texttt{encoder.6.output.dense}, with additional spikes in deeper encoder blocks. This corresponds to an even more catastrophic failure, with accuracy dropping from 91\% to only 1.25\%, underscoring the extreme vulnerability of transformers when critical dense and attention layers are compromised. The localization methodology thus identifies the most affected layers for targeted repair.
\begin{table}[h]
\centering
\captionsetup{labelfont=bf}
\caption{Layer Anomaly Scores for Attacked Models}
\label{tab:anomaly_scores}
\resizebox{\columnwidth}{!}{%
\begin{tabular}{lcc}
\hline
\textbf{Layer}          & \textbf{ResNet-18} & \textbf{BERT-Large} \\ \hline
Top Layer & \texttt{conv1} & \texttt{encoder.6.output.dense} \\
\textbf{Max Score}      & \textbf{50.000}    & \textbf{38.191}     \\
\textbf{Detection Rate} & \textbf{97.8\%}    & \textbf{74\%}       \\
\textbf{Accuracy Drop}  & \textbf{66.12\%}   & \textbf{89.75\%}    \\ \hline
\end{tabular}%
}
\end{table}

\subsubsection{\textbf{Attack Mitigation Performance}}
Table~\ref{tab:bitflip-results} evaluates robustness by comparing clean, post attack, and post mitigation accuracy using the z-score based repair of \BAREAI. Bit-flip attacks cause severe degradation; for example, ResNet-18 drops from 69.8\% to 0.1\%. \BAREAI~effectively restores CNN performance, recovering ResNet-50 to 72.5\% and VGG-16 to 69.8\%. The picture differs for language models: for LLaMA-3.2-1B with 100 bit flips, mitigation improves accuracy from 22\% to 27\%, and for BERT-Large recovery is limited to 9\% from a post attack accuracy of 1.25\%. The limited transformer recovery, despite detection accuracies of 74 to 84\%, is attributed to model depth: errors accumulate across the transformer stack during generation, and mean substitution cannot restore the precise values of repaired weights. In summary, detection accuracy exceeds 97\% for the CNN models and ranges from 74\% (BERT-Large) to 93\% (ViT-B) for transformer based models, while recovery is near clean for CNNs and ViTs and partial for deep language models. We report both outcomes to characterize where the current repair mechanism is and is not sufficient.
\begin{table}[h]
\centering
\captionsetup{labelfont=bf}
\caption{Model Accuracy Recovery Under Bit-Flip Attacks with \BAREAI~Mitigation}
\label{tab:bitflip-results}
\renewcommand{\arraystretch}{1.2}
\setlength{\tabcolsep}{2pt}
\begin{tabular}{@{}lccccc@{}}
\toprule
\textbf{Model} &
\textbf{\shortstack{Bit-Flips\\(\#)}} &
\textbf{\shortstack{Top-1 Acc.\\(Clean)}} &
\textbf{\shortstack{Acc. (After \\ Attack)}} &
\textbf{\shortstack{Acc. (After \\Mitigation)}} &
\textbf{\shortstack{Detection\\Accuracy}} \\
\midrule
AlexNet     & 10 & 83.4\% & 11.2\% & 81.7\% & 98.3\% \\
ResNet-18   & 15 & 69.8\% & 0.1\%  & 67.4\% & 97.8\% \\
ResNet-50   & 20 & 75.3\% & 5.2\%  & 72.5\% & 98.1\% \\
VGG-16      & 20 & 71.1\% & 3.9\%  & 69.8\% & 98.7\% \\
ViT-B       & 20 & 85.0\% & 25.0\% & 81.0\% & 93.0\% \\
MobileNetV2 & 20 & 78.0\% & 20.0\% & 74.0\% & 91.0\% \\
BERT-Large & 200 & 91.0\% & 1.25\% & 9\% & 74.0\% \\
Llama-3.2-1B & 100 & 67.3\% & 22.0\% & 27.0\% & 82.0\% \\
Qwen3-0.6B & 100 & 67.8\% & 15.0\% & 28.0\% & 84.0\% \\
\bottomrule
\end{tabular}
\end{table}

\begin{table}[h]
\centering
\captionsetup{labelfont=bf}
\caption{Detection and Recovery Accuracy under Large-Scale Bit-Flip Scenarios}
\renewcommand{\arraystretch}{1.3}
\label{tab:large_scale_faults}
\resizebox{\columnwidth}{!}{%
\begin{tabular}{lcccc}
\hline
\textbf{Scenario} & \textbf{Flips} & \textbf{Det. Acc. (\%)} & \textbf{Rec. Acc. (\%)} & \textbf{FAR (\%)} \\ \hline
\textbf{Random (AlexNet)}  & 1000 & 92.8 & 88.1 & 3.9 \\
\textbf{Random (ResNet-18)} & 1000 & 93.2 & 89.4 & 3.6 \\
\textbf{Adversarial (ViT-B)} & 100  & 93.4 & 90.7 & 3.8 \\ \hline
\cite{chen2021low} Baseline & 1000 & 78.5 & 74.9 & 5.2 \\
\cite{zhan2021improving}Baseline & 1000 & 82.3 & 76.2 & 4.7 \\ \hline
\end{tabular}
}
\vspace{1mm}
\end{table}

\begin{table}[t]
\centering
\captionsetup{labelfont=bf}
\caption{Detection, recovery, and post-attack accuracy of
\BAREAI~across vision and language models.}
\label{tab:quantization_merged_attack}

\scriptsize
\setlength{\tabcolsep}{2.5pt}
\renewcommand{\arraystretch}{0.90}

\resizebox{\columnwidth}{!}{%
\begin{tabular}{@{}lccccc@{}}
\toprule
\textbf{Model / Task} &
\textbf{Prec.} &
\shortstack{\textbf{Attack}\\\textbf{Acc. (\%)}} &
\shortstack{\textbf{Detect.}\\\textbf{Acc. (\%)}} &
\shortstack{\textbf{Recover.}\\\textbf{Acc. (\%)}} &
\shortstack{\textbf{Ovh.}\\\textbf{(\%)}} \\
\midrule

\multicolumn{6}{c}{\textbf{Vision Models}} \\
\midrule
\multirow{2}{*}{AlexNet (CIFAR-10)}
 & FP32 & 35.4 & 97.8 & 93.5 & 2.1 \\
 & INT8 & 33.1 & 96.2 & 91.8 & 2.0 \\

\multirow{2}{*}{ResNet-18 (CIFAR-10)}
 & FP32 & 29.3 & 98.1 & 94.1 & 2.3 \\
 & INT8 & 27.6 & 96.9 & 92.7 & 2.2 \\

\multirow{2}{*}{ViT-B (ImageNet)}
 & FP32 & 24.9 & 97.5 & 93.8 & 2.4 \\
 & INT8 & 22.7 & 95.4 & 91.2 & 2.3 \\

\midrule
\multicolumn{6}{c}{\textbf{Language Models: SST-2}} \\
\midrule
\multirow{2}{*}{LLaMA-3.2-1B}
 & FP32 & 21.5 & 82.4 & 27.6 & 2.9 \\
 & INT8 & 20.8 & 80.6 & 26.9 & 2.8 \\

\multirow{2}{*}{Qwen3-0.6B}
 & FP32 & 14.8 & 84.1 & 28.2 & 3.0 \\
 & INT8 & 15.2 & 83.1 & 28.7 & 2.9 \\

\midrule
\multicolumn{6}{c}{\textbf{Language Models: MMLU}} \\
\midrule
Qwen3-0.6B     & INT8 & 20.0 & 95.0 & 15.0 & 3.0 \\
LLaMA-3.2-1B  & INT8 & 35.0 & 90.0 & 12.0 & 3.0 \\
\bottomrule
\end{tabular}%
}

\vspace{-2mm}
\end{table}

\begin{table*}[h]
\centering
\captionsetup{labelfont=bf}
\caption{Impact of Magnitude Based Bit-Flip Attacks and Mitigation Effectiveness}
\label{tab:magnitude_bitflip}
\begin{tabular}{lcccccc}
\hline
\textbf{Model} & \textbf{Precision} & \textbf{\# Flips} &
\textbf{Clean Acc. (\%)} &
\textbf{After Attack (\%)} &
\textbf{After Mitigation (\%)} &
\textbf{Detection Acc. (\%)} \\
\hline
ResNet-18 (CIFAR-10) & FP32 & 50  & 93.1 & 41.7 & 58.4 & 96.8 \\
ResNet-18 (CIFAR-10) & FP32 & 100 & 93.1 & 29.3 & 46.9 & 97.2 \\

ViT-B (ImageNet) & FP32 & 50  & 81.2 & 38.6 & 52.7 & 95.9 \\
ViT-B (ImageNet) & FP32 & 100 & 81.2 & 24.9 & 41.8 & 96.3 \\

LLaMA-3.2-1B (SST-2) & FP32 & 100 & 67.3 & 21.5 & 27.6 & 82.4 \\
Qwen3-0.6B (SST-2)   & FP32 & 100 & 67.8 & 14.8 & 28.2 & 84.1 \\
\hline
\end{tabular}
\end{table*}

\textbf{Adversarial Multiple Bit Flip Scenarios:}
Evaluating larger and more diverse fault intensities provides a more convincing demonstration of scalability and robustness.
Accordingly, we extended our experiments to (i) \textit{adversarial multiple bit attacks} with up to 100 targeted bit flips across sensitive layers, and (ii) \textit{random large scale corruption} with up to 1000 random bit flips uniformly distributed over the model, across AlexNet, ResNet-18, and ViT-B. In Table~\ref{tab:large_scale_faults}, the results show more than 92\% detection and 88\% recovery accuracy with below 4\% false alarms. In comparison, the range restricted~\cite{chen2021low} and boundary aware~\cite{zhan2021improving} methods show 10 to 15\% higher false negatives and degrade beyond 200 flips. The runtime, model agnostic APC based design of \BAREAI~therefore sustains high resilience across architectures and fault intensities, making it suitable for long lifetime, safety critical AI accelerators.

\textbf{Robustness under Quantization:}
Robustness under quantization is critical for deploying \BAREAI~on edge accelerators. We evaluated \BAREAI~on 8-bit post-training quantized (PTQ) and quantization aware trained (QAT) versions of AlexNet, ResNet-18, and ViT-B. As shown in Table~\ref{tab:quantization_merged_attack}, \BAREAI~achieves above 95\% detection accuracy and above 89\% recovery accuracy under INT8 inference, within 2.3\% of FP32 performance. APC features (mean, sparsity, entropy, and higher moments) remain stable under quantization noise, while precision aware z-score normalization enables effective mitigation.
Table~\ref{tab:quantization_merged_attack} summarizes post attack accuracy, detection accuracy, recovery, and hardware overhead across vision and language models under FP32 and INT8 inference. For vision models, targeted attacks severely degrade accuracy (e.g., to 29.3\% for FP32 ResNet-18 and 24.9\% for FP32 ViT-B), with comparable vulnerability under INT8. \BAREAI~detects these attacks with above 95\% accuracy and restores above 91\% accuracy for INT8 ResNet-18 and ViT-B, demonstrating the robustness of APC based signatures under quantization.
For SST-2 language models, targeted bit flips reduce FP32 accuracy to 21.5\% for LLaMA-3.2-1B and 14.8\% for Qwen3-0.6B, with similar degradation under INT8. \BAREAI~detects these attacks with 80.6 to 84.1\% accuracy and recovers 5 to 13\% absolute performance via localized mitigation; as discussed above, the limited recovery stems from perturbation accumulation across deep transformer stacks.

In summary, \BAREAI~maintains high detection accuracy in both FP32 and INT8 settings with below 3\% hardware energy overhead, achieving near clean recovery for vision models and partial recovery for language models.

\subsubsection{\textbf{Impact of Magnitude-Based (DNL) Bit-Flip Attacks and Mitigation Effectiveness}}
Table~\ref{tab:magnitude_bitflip} evaluates \BAREAI~under magnitude based bit-flip attacks using the Deep Neural Lesion (DNL) strategy. This pass free attack flips sign bits of high magnitude weights without requiring data, gradients, or iterative optimization~\cite{galil2025no}, making it a realistic threat to deployed models. Even a small number of flips causes severe degradation: ResNet-18 drops from 93.1\% to 41.7\% with 50 flips and to 29.3\% with 100 flips, while ViT-B falls from 81.2\% to 24.9\% under 100 flips, demonstrating the effectiveness of magnitude based attacks across CNNs and transformers.
Despite this, \BAREAI~achieves above 95\% detection accuracy across all vision models, highlighting the effectiveness of APC based behavioral monitoring in identifying sign flip induced anomalies. Z-score based mitigation applied to localized faulty layers substantially restores accuracy, improving ResNet-18 from 29.3\% to 46.9\% and ViT-B from 24.9\% to 41.8\% under 100 bit attacks; note that sign flipped high magnitude weights are exactly the extreme outliers that the z-score filter is designed to catch, though mean substitution recovers regularity rather than the original values, which bounds the achievable restoration.

For large language models, perturbation accumulation across deep transformer stacks makes DNL attacks especially damaging. In LLaMA-3.2-1B and Qwen3-0.6B, 100 sign bit flips reduce accuracy to 21.5\% and 14.8\%, respectively. \BAREAI~detects these attacks with above 82\% accuracy and recovers up to 13\% absolute accuracy, preventing complete collapse but falling short of full restoration.
Overall, Table~\ref{tab:magnitude_bitflip} confirms that \BAREAI~remains effective at detecting pass free magnitude based DNL, gradient guided, and random bit-flip attacks, with recovery effectiveness that is strong for vision models and partial for LLMs.

\subsection{\textbf{Hardware Overhead for \BAREAI}}
\label{sec:hw_overhead}
In this section, we present the timing, area, and energy overhead incurred by integrating \BAREAI~into an AI accelerator. The analysis covers the hardware setup, area overhead, and the resulting timing and energy implications.

\begin{table*}[h]
\fontsize{10pt}{10pt}\selectfont
\renewcommand{\arraystretch}{1.1}
\centering
\captionsetup{labelfont=bf}
\caption{\BAREAI~framework incurs modest overhead (Energy, Area and Timing) across different neural network models}
\label{tab:probiotic_vs_at}
\resizebox{\textwidth}{!}{%
\begin{tabular}{cccccccccccc}
\hline
\multicolumn{4}{c}{\textbf{Baseline Information}} &
  \multicolumn{8}{c}{\textbf{\begin{tabular}[c]{@{}c@{}}Additional Energy, Area and Inference time requirement \\ for \BAREAI~Framework\end{tabular}}} \\ \hline
\multirow{2}{*}{\textbf{Network}} &
  \multirow{2}{*}{\textbf{\begin{tabular}[c]{@{}c@{}}Energy \\ (mJoules)\end{tabular}}} &
  \multirow{2}{*}{\textbf{\begin{tabular}[c]{@{}c@{}}Area \\ (Sq.mm)\end{tabular}}} &
  \multicolumn{1}{c}{\multirow{2}{*}{\textbf{\begin{tabular}[c]{@{}c@{}}Inference \\ time (s)\end{tabular}}}} &
  \multirow{2}{*}{\textbf{\begin{tabular}[c]{@{}c@{}}Energy \\ (mJoules)\end{tabular}}} &
  \multicolumn{3}{c}{\textbf{Area (mm²)}} &
  \multirow{2}{*}{\textbf{\begin{tabular}[c]{@{}c@{}}Inference \\ time (s)\end{tabular}}} &
  \multirow{2}{*}{\textbf{\begin{tabular}[c]{@{}c@{}}Energy \\ Overhead \\ (\%)\end{tabular}}} &
  \multirow{2}{*}{\textbf{\begin{tabular}[c]{@{}c@{}}Area \\ Overhead\\ (\%)\end{tabular}}} &
  \multirow{2}{*}{\textbf{\begin{tabular}[c]{@{}c@{}}Inference \\ time \\ Overhead \\ (\%)\end{tabular}}} \\ \cline{6-8}
 &
   &
   &
  \multicolumn{1}{c}{} &
   &
  \textbf{\begin{tabular}[c]{@{}c@{}}PULSE \\ Area \\ (mm²)\end{tabular}} &
  \textbf{\begin{tabular}[c]{@{}c@{}}APC \\ Submodule \\ Area (mm²)\end{tabular}} &
  \textbf{\begin{tabular}[c]{@{}c@{}}Total\\ Area \\ (Sq.mm)\end{tabular}} &
   &
   &
   &
   \\ \hline
Alexnet &
  28.70 &
  2.26 &
  0.62 &
  0.68 &
  0.043 &
  0.037 &
  0.08 &
  0.060 &
  2.37\% &
  3.50\% &
  9.68\% \\
ResNet-18 &
  71.30 &
  3.30 &
  0.8 &
  1.78 &
  0.043 &
  0.088 &
  0.131 &
  0.080 &
  2.51\% &
  3.87\% &
  9.94\% \\
ResNet-50 &
  150.50 &
  4.52 &
  1.2 &
  3.88 &
  0.043 &
  0.112 &
  0.155 &
  0.144 &
  2.58\% &
  3.43\% &
  11.94\% \\
VGG-16 &
  613.80 &
  6.78 &
  1.27 &
  15.22 &
  0.043 &
  0.217 &
  0.260 &
  0.118 &
  2.48\% &
  3.82\% &
  9.28\% \\ \hline
\end{tabular}%
}
\end{table*}

\noindent{\textbf{Hardware Setup:}}
Table~\ref{tab:probiotic_vs_at} summarizes the overhead associated with integrating the \BAREAI~framework into an AI accelerator. The unit synthesized here is the deployed on-chip realization described in Section III-A.2: a compact, dedicated ML engine comprising a fixed three layer neural network with a 16-8-4 neuron configuration that executes the distilled PULSE decision function on compressed statistical features (e.g., sparsity, entropy, zero count, layer wise MACs) extracted by the APCs. Because this engine is optimized for real time processing, it requires minimal logic and memory.
In addition, \BAREAI~incorporates model specific input buffers that include counters, configurable logic (e.g., thresholds for sparsity and entropy, FSM states), and controller adaptations tailored to the target DNN.

\noindent{\textbf{Area Overhead:}}
The PULSE engine and the complete \BAREAI~module (APC + PULSE + interface logic) are synthesized using Synopsys Design Compiler on a 28nm CMOS technology node. While the PULSE core logic remains constant across models (0.043mm\textsuperscript{2}), the APC submodule area grows for deeper networks such as ResNet-50 and VGG-16 due to additional counters, FSMs, and entropy and sparsity monitors. For instance, the APC area increases from 0.037mm\textsuperscript{2} in AlexNet to 0.217mm\textsuperscript{2} in VGG-16.

We adopt Eyeriss \cite{chen2016eyeriss} as the baseline accelerator due to its silicon validated architecture and relevance to edge AI, allowing a reliable estimation of the integration overhead of \BAREAI. The original Eyeriss chip, fabricated in 65nm CMOS for AlexNet inference, occupies approximately 12.25 mm\textsuperscript{2} on a 3.5 mm by 3.5 mm die \cite{chen2016eyeriss}. We use these measurements as validated references for area and energy normalization. For ResNet-18, ResNet-50, and VGG-16, we scaled the Eyeriss architecture according to the compute and memory requirements of each model. The area at 28nm is first estimated for AlexNet using quadratic scaling, followed by a sublinear MAC based scaling model with exponent $\alpha = 0.6$ to derive accelerator area for deeper networks, consistent with the methodology in \cite{han2016eie}.

In contrast, the proposed \BAREAI~(APC+PULSE) subsystem is not derived from Eyeriss silicon measurements; it is independently synthesized in the 28nm technology node. To integrate synthesized \BAREAI~with the DNN accelerator for a fair and consistent comparison, these estimates assume an Eyeriss architecture ported to 28nm without architectural modifications. Based on this approach, the scaled area estimates for popular CNN models are tabulated in Table~\ref{tab:probiotic_vs_at} as follows: AlexNet 2.27mm\textsuperscript{2}, ResNet-18 3.39mm\textsuperscript{2}, ResNet-50 4.52mm\textsuperscript{2}, and VGG-16 6.78mm\textsuperscript{2}. We note that these accelerator side figures are scaling based estimates rather than measured silicon, and that the \BAREAI~overhead percentages should be read relative to this estimation methodology.

\begin{table*}[htb]
\centering
\captionsetup{labelfont=bf}
\caption{Comparison of Existing Defenses Against Bit-Flip Attacks (BFAs)}
\label{tab:bfa_defenses}
\begin{tabular}{|p{3cm}|p{4.5cm}|p{5cm}|p{3cm}|}
\hline
\textbf{Defense Approach} & \textbf{Methodology} & \textbf{Merits} & \textbf{Limitations} \\ \hline
\textbf{Weight Reconstruction + ECC} \cite{li2020defending} &
Reconstructs corrupted weights using redundancy; integrates error-tolerant loss with ECC. &
Improves robustness against random and injected faults; effective recovery of weights. &
Requires costly retraining and prior fault identification. \\ \hline
\textbf{RA-BNN} \cite{siraj2021ra} &
Robustness-aware binary neural networks with weight clipping, gradient masking, and QAT. &
Enhances resilience without sacrificing accuracy; efficient for low-bit models. &
Limited to binary/quantized models; risks degraded generalization. \\ \hline
\textbf{Hybrid Sensitivity-Based Protection} \cite{wang2302aegis} &
Layer-wise sensitivity analysis to selectively protect critical weights. &
Balances robustness and efficiency; reduces overhead by targeting only critical parameters. &
Still vulnerable to adaptive and targeted attacks. \\ \hline
\textbf{Hardware ECC / SECDED\cite{schechter2010chipkill,zhou2023dnn_defender}} &
On-chip memory error correction (single/double error detection and correction). &
Transparent protection; effective against random faults. &
Ineffective against adversarially targeted flips; adds area, power, and latency overhead that grows with correction strength. \\ \hline
\textbf{Randomization (Address Scrambling, Weight Encoding)\cite{velcicky2024deepncode}} &
Obfuscates weight storage to hinder precise adversarial targeting. &
Lightweight; increases attacker effort. &
Only delays attacks, does not fully prevent them. \\ \hline
\textbf{Runtime Detection (NeuroPots)} \cite{liu2023neuropots} &
Detects and corrects bit-flip anomalies during inference using activation monitoring. &
Prevents accuracy collapse in real-time; no retraining needed. &
Detection overhead; scalability challenges for large models. \\ \hline
\textbf{Offline Recovery (DeepHammer, Impactful Search)} \cite{yao2020deephammer, benedek2024impactful, kundu2024bit} &
Attack-aware analysis with advanced bit-search algorithms. &
Exposes model vulnerabilities, enabling stronger defensive training. &
Not a defense itself; highlights attack surface. \\ \hline
\textbf{Proposed \BAREAI} &
Real-time framework leveraging APCs and PULSE to capture layer-wise signatures and detect tampering, with ASI localization and z-score repair. &
Runtime detection, localization, and repair without weight inspection or retraining; up to 98\% detection on vision models; monitoring cost independent of attack strength. &
Recovery is partial for deep LLMs; detection weakens on low-magnitude mantissa flips; relies on a trusted on-die monitor (Section III-A.3). \\ \hline
\end{tabular}
\end{table*}

\noindent{\textbf{Timing and Energy Overhead:}}
Integration of \BAREAI~leads to modest increases in inference latency and energy usage. Across all models, the average inference time increases by 10.21\%, while energy consumption grows by roughly 3\%. For example:
\begin{itemize}
    \item AlexNet: +9.69\% timing, +2.37\% energy
    \item ResNet-18: +9.94\% timing, +2.51\% energy
    \item ResNet-50: +11.94\% timing, +2.58\% energy
    \item VGG-16: +9.28\% timing, +2.48\% energy
\end{itemize}
These results indicate that while the timing cost is not negligible, it remains controlled and within practical limits, and can be reduced further through counter selection as discussed next.

\begin{figure}[t!]
    \centering
    \captionsetup{labelfont=bf}
    \includegraphics[width=0.85\linewidth]{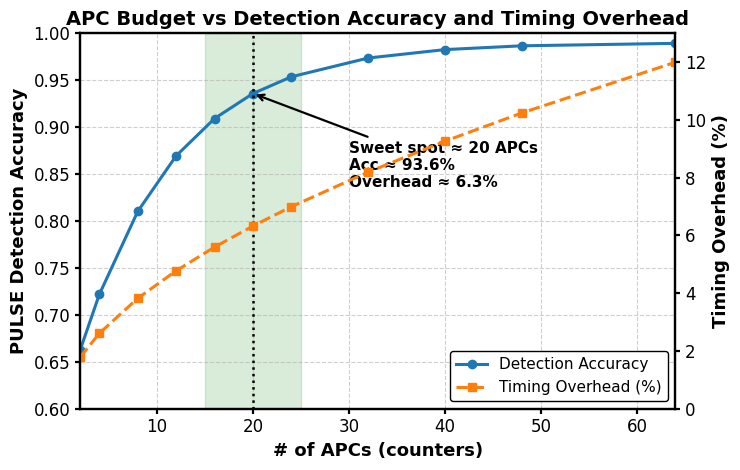}
    \caption{Trade-off between APC budget, detection accuracy, and timing overhead.
    }
    \label{fig:apc_budget}
\end{figure}
\vspace{-5mm}
\subsection{\textbf{APC Budget vs. Detection Accuracy and Overhead}}
Figure~\ref{fig:apc_budget} quantifies the trade-off between the number of APCs monitored, the resulting detection accuracy, and the timing overhead. Detection accuracy saturates beyond 20 counters, achieving 93.6\% accuracy with only 6.3\% timing overhead, which we identify as the preferred operating point. Beyond this point, additional counters yield marginal accuracy gains while incurring linear increases in overhead. This shows that \BAREAI~can achieve strong protection with a modest number of counters, ensuring practical deployability.
\BAREAI~is primarily targeted at edge class accelerators such as Eyeriss like inference cores and mobile NPUs, where the reported overhead (below 3\% energy, below 4\% area, approximately 10\% latency at the full counter budget) falls within typical design margins. The APC modules are fully configurable: a subset of APCs (e.g., monitoring only sparsity and entropy) can be selectively enabled to balance detection accuracy and overhead according to application constraints. As shown in Figure~\ref{fig:apc_budget}, detection accuracy saturates at approximately 20 counters with about 6\% timing overhead, and further downscaling can reduce the energy overhead to below 1\%.

\section{\textbf{Discussions}}
\label{sec:discussion}
In this section, we discuss how \BAREAI~compares with state of the art defenses. Table~\ref{tab:bfa_defenses} compares \BAREAI~with previous defenses based on their methodology, merits, and limitations. \BAREAI~avoids the model and precision limits of RA-BNN and sensitivity based protection, and it circumvents the expensive retraining and prior fault identification requirements of weight reconstruction with ECC by operating at inference time. By monitoring activation signatures through APCs and PULSE, \BAREAI~achieves up to 98\% detection on vision workloads with low accuracy loss, in contrast to hardware ECC/SECDED and randomization, which either miss or merely delay targeted flips. For real time responsiveness, \BAREAI~employs lightweight layer wise signatures with ensemble decisioning, in contrast to NeuroPots style runtime monitors that face overhead and scalability challenges for large models. It requires no changes to storage formats, is independent of model and precision, and can operate in tandem with ECC. At the same time, Table~\ref{tab:bfa_defenses} records the current limitations of our approach honestly: recovery is partial for deep language models, detection weakens for low magnitude mantissa flips, and the guarantees rest on the monitor security analysis of Section~\ref{sec:monitor_security}.

Although ECC has been the traditional approach for protecting memory against random soft errors, it faces fundamental limitations when applied to DNNs under bit-flip attacks. Adaptive adversaries can concentrate multiple flips within the same codeword, defeating single error correction and forcing the use of stronger multi-bit correction, which drives rapidly increasing redundancy in area, latency, and energy. Formally, consider a neural network with $k_{\text{total}}$ weight bits, of which $k_{\text{pro}}$ are protected using an $(n,k)$ linear code. Normalized redundancy (fractional overhead) is given by
\cite{huang2020functional}:
\begin{equation}
r = \frac{k_{\text{pro}}(n-k)}{k_{\text{total}} \, k},
\label{eq:ecc-overhead}
\end{equation}
where $k$ denotes the number of information bits in each codeword and $(n-k)$ the number of parity bits.

To tolerate up to $t$ adversarial bit flips per protected word, BCH type constructions require at most $m t$ parity bits, where $m \approx \log_2 n$. The redundancy thus scales approximately as:
\begin{equation}
r_{\text{BCH}} \approx \frac{k_{\text{pro}}}{k_{\text{total}}} \cdot \frac{t \, \log_2 n}{k}.
\label{eq:bch-overhead}
\end{equation}
This expression highlights that ECC overhead grows \textit{linearly with $t$}, the level of resiliency against multi-bit errors: protecting against more powerful attacks requires proportionally more hardware. Practical implementations additionally incur latency for syndrome computation and correction, along with interleaving and buffering to avoid clustered errors.

In contrast, \BAREAI~does not rely on storing redundant bits. It leverages layer wise APCs, such as sparsity, entropy, and cosine drift, to monitor abnormal deviations caused by bit-flip attacks. The overhead depends only on the number of layers ($L$) and counters per layer ($K$):
\begin{equation}
\text{Overhead}_{\BAREAI} = \alpha L K,
\label{eq:bareai-overhead}
\end{equation}
where $\alpha$ is a small implementation constant representing the cost of each counter. Crucially, this cost is \textit{independent of the number of adversarial flips $t$}. Whether the adversary flips 1 bit or 100 bits, the detection infrastructure remains constant.

ECC overhead grows as $\mathcal{O}(t)$ with rising resiliency requirements, while \BAREAI~incurs constant $\mathcal{O}(1)$ overhead regardless of attack strength. This scalability advantage makes \BAREAI~more efficient for protecting DNNs against adaptive bit-flip attacks, which often involve clustered multi-bit flips that are especially costly for ECC to handle. The same asymmetry underlies the monitor hardening argument of Section~\ref{sec:monitor_security}: the small, constant size of the trusted monitoring base is what makes it cheap to protect.

\section{\textbf{Conclusion}}
We presented \BAREAI, a low cost runtime defense against bit-flip attacks in DNN accelerators. By leveraging embedded AI Performance Counters and the lightweight PULSE detector, \BAREAI~detects and localizes adversarial faults with up to 98\% accuracy on vision models and between 74\% and 95\% on language models, while incurring modest energy (below 3\%), area (below 4\%), and latency (approximately 10\%, reducible to about 6\%) overhead. The framework captures subtle runtime deviations induced by random, targeted, adaptive, and magnitude based bit flips across CNNs, Vision Transformers, and LLMs. Unlike ECC or retraining based defenses, \BAREAI~is attack agnostic and integration friendly, provides actionable insight through the ASI, enables statistical repair of corrupted layers through z-score based restoration, and maintains a monitoring cost that is independent of attack strength. Recovery is near clean for CNNs and ViTs and currently partial for deep language models, where errors accumulate across the transformer stack. Future work will address stronger repair mechanisms for LLMs, including per-channel and per-head statistics, silicon or FPGA level validation of the APC extraction pipeline, robustness characterization under benign distribution shift, and formal side channel evaluation of the monitoring infrastructure.

\section*{Acknowledgment}
This work was supported in part by the National Science Foundation under Grants 2537757 and 2350365; the Semiconductor Research Corporation under Task 3243.001; and the Purdue Center for Secure Microelectronics Ecosystem under Grant CSME\#210205.
\vspace{-5pt}
{\footnotesize
\bibliographystyle{IEEEtran}
\balance
\bibliography{bibliography}
}

\end{document}